\documentclass[%
 reprint,
 twocolumn,
 amsmath,amssymb,
 aps,
 prc,
 floatfix,
]{revtex4}

\usepackage{graphicx}% Include figure files
\usepackage{caption}
\usepackage{subcaption}

\usepackage{dcolumn}% Align table columns on decimal point
\usepackage{bm}% bold math
\usepackage{xcolor}

\newcommand{\mref}[1]{\textbf{\textsc{\textcolor{red}{[REF]}}}}

\newcommand{\be}{\begin{equation}}
\newcommand{\ee}{\end{equation}}

\begin{document}

\title{Deuteron, triton, helium-3 and hypertriton production in relativistic heavy-ion collisions via stochastic multi-particle reactions}

\author{M. Ege$^{1,3}$} \email{mege@itp.uni-frankfurt.de}

\author{J. Mohs$^{1,3}$} \email{jmohs@itp.uni-frankfurt.de}

\author{J. Staudenmaier$^{1}$\footnote{left academia}}

\author{H. Elfner$^{2,1,3,4}$}
\email{h.elfner@gsi.de}

\address{1 Institute for Theoretical Physics, Goethe University, Max-von-Laue-Strasse 1, 60438 Frankfurt am Main, Germany}
\address{2 GSI Helmholtzzentrum f\"ur Schwerionenforschung, Planckstr. 1, 64291 Darmstadt, Germany}
\address{3 Frankfurt Institute for Advanced Studies, Ruth-Moufang-Strasse 1, 60438 Frankfurt am Main, Germany}
\address{4 Helmholtz Research Academy Hesse for FAIR (HFHF), GSI Helmholtz Center, Campus Frankfurt, Max-von-Laue-Straße 12, 60438 Frankfurt am Main, Germany}

\date{\today}

\begin{abstract}
The production of light nuclei in heavy-ion collisions is an excellent probe for studying the phase diagram of quantum chromodynamics and for the search of a critical end point.
In this work we apply a hybrid approach in which we study the light nuclei production in the afterburner stage of central Au+Au collisions at $\sqrt{s}_{NN}=7.7$, 14.5 and 19.6 GeV.
In this stage, light nuclei are produced dynamically in $4\leftrightarrow 2$ catalysis reactions.
A comparison of  the dynamic production and a coalescence approach is presented for transverse momentum spectra of deuterons, tritons, $^3\rm He$ nuclei and hypertritons and ratios of light nuclei yields.
A good agreement with the experimentally measured yield of nuclei is found and we proceed to further investigate the production mechanisms of light nuclei by calculating the rates of the important channels for the formation and disintegration.
We find that the afterburner stage is essential for the description of light nuclei formation in heavy-ion collisions, as light nuclei undergo a large number of interactions. 
\end{abstract}

\maketitle

% \tableofcontents

%#######################################################
\section{Introduction}
%#######################################################

Heavy-ion collisions provide a unique opportunity to study strongly-interacting matter under extreme conditions under controlled conditions in a laboratory.
The structure of the phase diagram of quantum chromodynamics (QCD) is of current interest and many experimental efforts with the aim to shed light on the phase structure are carried out, for example within the Beam Energy Scan II (BES II) program at the Relativistic Heavy Ion Collider (RHIC) \cite{STAR:2014}.

Light nuclei production in heavy-ion collisions can serve as a probe of hot and dense QCD matter.
Understanding how light nuclei with binding energies of only a few MeV can be produced in a hot medium is a topic of current research.
The deuteron yield as measured by the ALICE collaboration \cite{Adam:2019wnb} can be described with the statistical hadronisation model \cite{Andronic:2010qu,Andronic:2017pug}.
This suggests that light nuclei are formed and survive at a temperature of $\sim 150$ MeV.
Studies within afterburner calculations have shown that this does not necessarily correspond to the chemical freeze-out of deuterons but their yield stays constant as formation and disintegration processes are balanced \cite{Oliinychenko:2018ugs}.
A similar result was obtained by solving the coupled reaction rate equations for the light nuclei yields in an expanding system \cite{Neidig:2021bal}.

The possible critical point is one of the most interesting features of the QCD phase diagram.
The production of light nuclei was suggested to serve as a probe for critical fluctuations \cite{Bzdak:2019pkr}.
Calculations within a coalescence model allow to relate fluctuations and correlations to the light nuclei ratios \cite{Sun:2017xrx,Sun:2018jhg}.
The STAR collaboration hence published the measured particle number ratios $N_d/N_p$, $N_t/N_p$ and $N_tN_p/N_d^2$ \cite{STAR:2022hbp}.
Further microscopic calculations allow to relate light nuclei ratios to the interaction between nucleons near the critical point \cite{Shuryak:2018lgd,Shuryak:2019ikv}.

The production of deuterons and other light nuclei can be studied from a theory perspective in various approaches.
Many works successfully apply a transport description for the evolution of a heavy-ion collision.
One possibility to produce light nuclei based on a transport model is to apply coalescence based on the distribution of nucleons in phase-space in the final state of the transport calculation.
Coalescence is the most extreme scenario of late stage formation of light nuclei.
Deuteron formation with a coalescence approach applied to UrQMD \cite{Bass:1998ca} calculations has been studied for a wide range of energies in \cite{Sombun:2018yqh}.
A further study within the UrQMD model takes more light nuclei into account \cite{Deng:2020zxo} and the aforementioned particle ratios are presented as a function of collision energy.
A nontrivial structure in the excitation function is observed but is not related to critical fluctuations as they are not incorporated in the model.
Light nuclei production via coalescence has also been studied in the JAM model \cite{Nara:1999dz} and the particle ratios are shown as a function of the collision energy as well \cite{Liu:2019nii}.
In this calculation, hardly any energy dependence of the $N_tN_p/N_d^2$ ratios is found.

Another possibility to realize light nuclei production in a transport model is via scatterings.
This approach was applied in the aforementioned work \cite{Oliinychenko:2018ugs} to describe the production of deuterons at the LHC within the SMASH transport model \cite{Weil:2016zrk}.
This work was continued for lower collision energies to compare to data from the RHIC BES \cite{Oliinychenko:2020znl}.
The flow of deuterons within this model has been studied at SIS18 energies \cite{Mohs:2020awg}.
While in these works the deuterons are formed in binary reactions, including an artificial intermediate state for the pion and nucleon catalysis reactions, the stochastic collision criterion was applied in a further step to perform multi-particle reactions \cite{Staudenmaier:2021lrg}.
Light nuclei production via multi-particle reactions has been studied first in \cite{Danielewicz:1991dh}.
The relativistic kinetic equations from this work have been recently applied to study the production of deuterons, tritons and $^3\mathrm{He}$ nuclei as well as yield ratios at RHIC and LHC energies \cite{Sun:2022xjr}.
In \cite{Coci:2023daq} multi-particle reactions are implemented to investigate deuteron formation within the PHQMD model \cite{Aichelin:2019tnk}.
In the PHQMD framework clusters can be identified by using the minimum spanning tree procedure \cite{Aichelin:1991xy} or the Simulated Annealing Clusterization Algorithm\cite{Puri:1996qv,Puri:1998te}.

In this work, we investigate the dynamic production of deuterons, tritons, $^3\mathrm{He}$ nuclei and hypertritons in multi-particle reactions in central heavy-ion collisions at BES II energies.
For this purpose, we extend the model used in \cite{Staudenmaier:2021lrg}, where multi-particle reactions are applied for the production of deuterons, by implementing the $4\leftrightarrow 2$ reactions to form $A=3$ nuclei in SMASH.
Like in the previous work, light nuclei are formed in the afterburner stage of a hybrid approach, where the MUSIC code \cite{Schenke:2010nt,Schenke:2010rr,Paquet:2015lta} is applied to solve the (3+1)D viscous hydrodynamic equations. 
Calculations of multiplicities, transverse momentum spectra and particle ratios are confronted with experimental data from the STAR collaboration and a good agreement is found.
Further we explore how nuclei are formed by investigating the time evolution of the multiplicities of nuclei and the impact of all relevant reaction channels.
We observe that many reactions involving light nuclei take place in the afterburner stage from which we conclude that this stage is essential for a realistic description of light nuclei production in heavy-ion collisions.
A comparison between the dynamic formation of light nuclei and a coalescence approach is provided and both approaches yield a good description of the measurements for the transverse momentum spectra and the single ratios of the multiplicities.
The coalescence approach performs slightly better for the $N_tN_p/N_d^2$ ratio without any critical behaviour in the model.

Regarding the structure of this work, we begin by briefly describing the transport model SMASH and the hybrid approach used for this work in Section \ref{sec:model}.
We continue in Section \ref{sec:multiparticle} with the cross sections for the production of light nuclei and expound the realisation of the $4\leftrightarrow 2$ scatterings in our model.
Subsequently, details on the coalescence model are given in Section \ref{sec_coalescence}.
Before advancing to the results, the multi-particle reactions are validated in Section \ref{sec:validation} by comparing to rate equations in a box calculation.
The results section begins with the transverse momentum spectra in Section \ref{sec:transverse_momentum} and the particle ratios in Section \ref{sec:ratios} which are compared to experimental data and coalescence calculations.
We proceed by presenting the time evolution of light nuclei multiplicities in Section \ref{sec:multiplicities} and close the result section with a study of the reaction rates in Section \ref{sec:productionmech}.
Finally, we briefly summarise our findings and draw conclusions in Section \ref{sec:conclusions}.

%#######################################################
\section{Model Description \label{sec:model}}
%#######################################################

This work aims to resolve shed further light on the formation of light nuclei in heavy-ion collisions.
For this purpose, a dynamic model is essential as it allows to access the time evolution of the system.
Hybrid approaches have proven to be well suited for the description of heavy-ion collisions in the considered energy region.
We employ a hybrid model that was previously developed in \cite{Oliinychenko:2020znl} for a study of deuteron production and further applied in \cite{Staudenmaier:2021lrg}.

The evolution of the hot and dense phase of a heavy-ion collision can be formulated with the (3+1)D relativistic viscous hydrodynamic equations.
A solution of the hydrodynamic equations is obtained using the MUSIC code \cite{Schenke:2010nt,Schenke:2010rr,Paquet:2015lta,Denicol:2018wdp} where the equation of state from \cite{Monnai:2019hkn} is used.
The evolution starts after an initial state from a collision-geometry-based 3D model \cite{Shen:2020jwv}.
As the fireball expands, the system becomes more dilute and is more realistically described within a transport model.
As in the previous work, an energy density of $\epsilon=0.26$GeV/fm$^{3}$ is used to define the hypersurface where the switch between models takes place.
The hypersurface is found using CORNELIUS \cite{Huovinen:2012is}.
The switch from a macroscopic to a microscopic description is performed using Cooper-Frye sampling \cite{Cooper:1974mv}.
For the final stage, we incorporate the SMASH transport model \cite{Weil:2016zrk} for hadronic rescattering.

SMASH is based on the relativistic Boltzmann equation in which the system is described in terms of hadronic degrees of freedom.
Hadrons can undergo elastic scatterings or interact inelastically in the model by forming resonances which can further decay and thereby produce new particles.
The properties of the hadrons and resonances are adopted from the Particle Data Group \cite{ParticleDataGroup:2018ovx}.
Vacuum Breit-Wigner functions are assumed for the spectral functions of resonances.
A further possibility for hadrons to interact is via string excitation and fragmentation.
The implementation of the string model is described in \cite{Mohs:2019iee}.
Nuclear potentials are implemented in the model but are neglected here as their influence decreases at higher collision energies.

The production of light nuclei can be performed in two possible ways.
A dynamic description of the light nuclei formation within catalysis reactions is confronted with a coalescence approach.
The treatment for the multi-particle reactions required for light nuclei formation via pion and nucleon catalysis is described in the following and subsequently the coalescence approach is presented.

\subsection{Multi-particle reactions for $A= 3$ nuclei\label{sec:multiparticle}}

\begin{figure}
    \centering
    \includegraphics[width=\columnwidth]{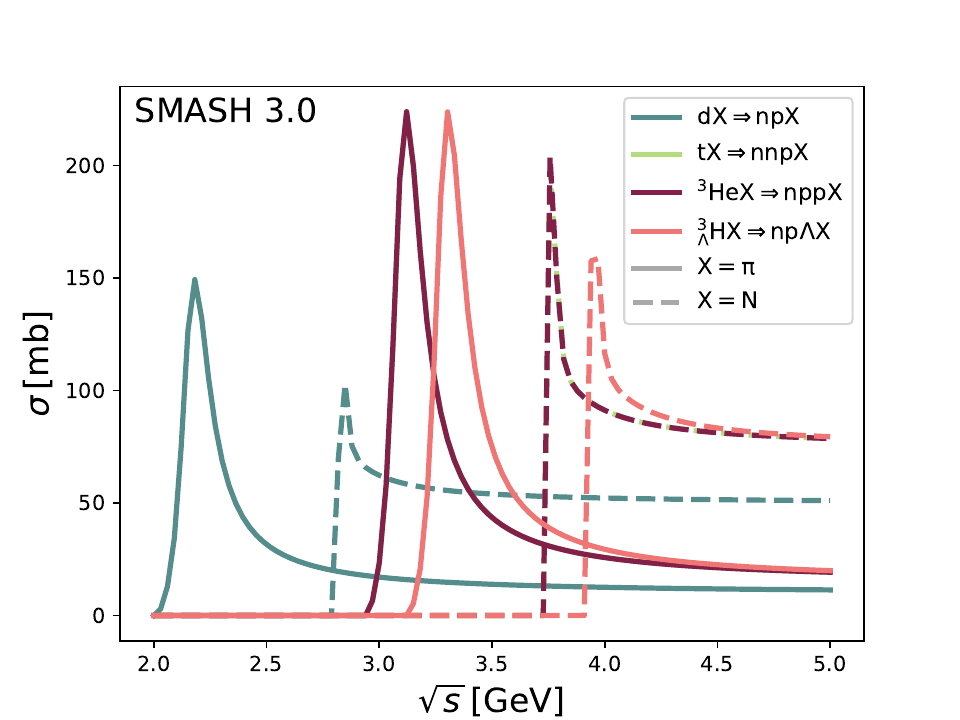}
    \caption{Interaction cross-sections for $2\rightarrow 3$ and $2\rightarrow 4$ light nuclei catalysis reactions as a function of the center of mass energy in SMASH.
    The full lines depict the cross sections for interactions with pions whereas the dashed lines represent interactions with nucleons.}
    \label{fig:cross}
\end{figure}

Multi-particle reactions to investigate the production of $d$, $t$, $He^3$ and $H^3_L$ are realized by \emph{stochastic rates}.
In contrast to geometric collision treatments, the stochastic treatment is straight-forward to extend to non-binary reactions.
The treatment is based on the collision probability, which is directly derived from the collision integral of the Boltzmann equation.
It is defined for a given time-step $\Delta t$ and a sub-volume $\Delta^3 x$.
The collision probability is used for a Monte-Carlo decision whether particles interact.
This is done for all possible combinations of particles in a given cell.
This method relies on a sufficient number of test particles per cell.
For this purpose each particle is represented by multiple test particles.
The details on the here employed stochastic collision treatment are given in \cite{Staudenmaier:2021lrg}, which also introduces its usage for the deuteron catalysis reaction.
This work is an extension of the same idea, i.e. describing the light nuclei formation via multi-particle catalysis reactions.

The reactions newly-introduced in this work to produce light nuclei are $Xnnp \leftrightarrow Xt$, $Xnpp \leftrightarrow XHe^3$ and $Xnp\Lambda \leftrightarrow XH^3_L$ with the catalyst $X = {\pi,N}$.
The $d$ reactions ($Xnp \leftrightarrow Xd$) from \cite{Staudenmaier:2021lrg} are also included in the presented calculations.

The collision probability for the new 4-to-2 reactions that produce $t$, $He^3$ or $H^3_L$ is
\begin{align}
    P_{4\rightarrow 2} =\left( \frac{g_{1'} g_{2'}}{g_1 g_2 g_3g_4} \right)\frac{\cal{S}!}{\cal{S}'!} \frac{1}{16E_1E_2E_3E_4} \frac{\Delta t}{(\Delta^3x)^3} \\ \nonumber
    \times \frac{\lambda(s;m^2_{1'},m^2_{2'})}{\Phi_4}\frac{\sigma_{2\rightarrow 4}}{4\pi s}
\end{align}
with the spin degeneracy factors $g$, the symmetry factor for indistinguishable states $\cal{S}$, the 4-body phase space $\Phi_4$ and the so called K\"all\'en function defined as $\lambda(s;m^2_{1'},m^2_{2'})=(s-m^2_{1'}-m^2_{2'})^2-4m^2_{1'}m^2_{2'}$.
Primed quantities belonging to the final and non-primed quantities belonging to the initial state.
The notation matches the previous works employing the method ~\cite{Staudenmaier:2021lrg, Garcia-Montero:2021haa}.
The cross-section of the reverse process is denoted by $\sigma_{2\rightarrow 4}$ (see below for details).
The expression of the 4-body collision probability in terms of the cross-section of the reverse process is based on the assumption that the scattering matrix element of the process only depends on the initial center-of-mass energy.

The probability for the 2-to-4 nuclei dissociation is the same as for all 2-body reactions (cf. \cite{Staudenmaier:2021lrg}, Eq. (4))
\begin{equation}
    P_{2 \rightarrow 4} = \frac{\Delta t}{\Delta^3 x} v_{\textrm{rel}} \sigma_{2\rightarrow 4} (\sqrt{s})
\end{equation}
with the abbreviation for the relative velocity $v_{\textrm{rel}}= \lambda^{1/2}(s;m_1^2,m_2^2)/({2E_1 E_2})$.

The cross section for the $2\rightarrow 4$ process is obtained by an extrapolation of the proton-deuteron and pion-deuteron cross sections.
For this purpose, the crosss sections between the deuteron and the catalysts are expressed in terms of the kinetic energy  
\begin{equation}
    T=\frac{(s-m_{A}+m_{\mathrm{cat}})^2}{2m_{A}}\,.
\end{equation}
The mass of the nucleus and the mass of the catalyst are denoted by $m_A$ and $m_\mathrm{cat}$.
In a simple picture, we consider only the number of constiuents of the nuclei such that the cross section of a  $A=3$ nucleus is equal to $3/2$ times the cross section of a deuteron at the same kinetic energy.
To summarize, the cross section for all $A=3$ nuclei with a given catalyst $X$ is 
\begin{equation}
    \sigma_{AX\rightarrow NNNX}(T) = \frac{A}{2}\sigma_{dX\rightarrow pnX}(T)\,.
\end{equation}
The cross sections are given in terms of $\sqrt{s}$ in Figure \ref{fig:cross} for both nucleon and pion catalysis reactions.
The difference between the cross sections for different nuclei emerge from the different masses as the kinetic energy at a given center of mass energy is smaller for heavier particles.
Since the triton and $^3\rm He$ nuclei have very similar masses, the curves overlap in Figure \ref{fig:cross}.
The angular distributions are treated isotropically.
We would like to emphasize that the large cross sections do not imply an interaction over a large distance in the transport description, since collision finding is based on small cells as described above.
Even though light nuclei are large sized objects, they are represented as point-like particles in the test particle picture as all other particles in SMASH.

\subsection{Production via coalescence}
\label{sec_coalescence}
An alternate way of modeling the production of light nuclei is by final state coalescence.
In this scenario, we model the time evolution of a heavy-ion collision without taking light nuclei into account as active degrees of freedom.
Instead, the light nuclei are produced in the final state of the collision, based on the phase-space distribution of protons and neutrons.
As multi-particle reactions are not required in this setup, we apply the geometric collision criterion in this case which does not rely on many test particles.
Using only a single test particle per real particle, coalescence can be carried out with physical coalescence parameters without rescaling to account for the oversampling.
We take the distance in both configuration and momentum space into account for deciding whether coalescence between nucleons takes place.
As the particles drift further apart as they are propagated until the end time of the calculation, which is a technical parameter, we fix the time for coalescence for each pair of particles individually to their last interaction time as it was done in \cite{Sombun:2018yqh}.
We further boost to the two-particle rest-frame to evaluate the difference in momentum and position and compare them to the cutoff parameters $\Delta p = 300\,\rm MeV$ and $\Delta r = 3\,\rm fm$ respectively.
These parameters were chosen to roughly reproduce light nuclei yields at the HADES experiment for an upcoming work \cite{Mohs:1995npc}.
Coalescence is allowed if both are smaller than the cutoff parameters.
We apply the same mechanism to form various light nuclei species but for simplicity keep the coalescence parameters the same.
The coalescence procedure starts by identifying deuterons from pairs of protons and neutrons.
In further steps larger nuclei are produced by combining existing nuclei with protons, neutrons, $\Lambda$ baryons or other nuclei based on the aforementioned coalescence criterion.
More details are provided in \cite{Mohs:1995npc}.
Critical fluctuations are not considered in the coalescence model.

%#######################################################
\section{Validation\label{sec:validation}}
%#######################################################

Before showing results in the next section, we put the newly introduced multi-particle reactions to the test.
For validation a comparison to an analytical solution is well suited.
An analytical solution for particle multiplicities in a box with a limited number of processes can be obtained by solving the rate equations \cite{Pan:2014caa}.
Note that the description of a realistic heavy-ion collision requires the full transport calculation as it incorporates all non-equilibrium dynamics and numerous cross sections.
We extended the rate equations used in \cite{Staudenmaier:2021lrg} to further include the reactions for triton, helium and hypertriton.
The equations are given in Appendix \ref{ap_rate_equations}.

\begin{figure}
    \centering
    \includegraphics[width=\columnwidth]{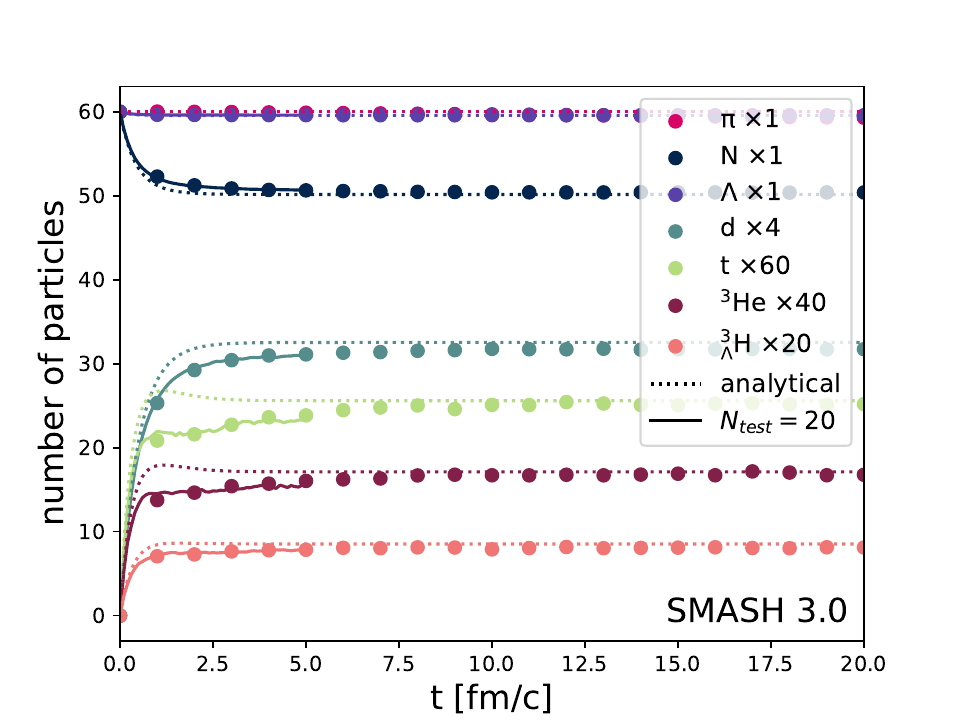}
    \caption{Particle multiplicities over time in a box of $(10\text{fm})^{3}$ and $T=150$ MeV. The circles represent calculations with 10 test particles and the full lines represent calculations performed with 20 test particles.
    The solutions from the rate equations are shown as dotted curves.}
    \label{fig:boxtest}
\end{figure}

The results from the rate equation are compared with the multiplicities from the SMASH calculation in Figure \ref{fig:boxtest}.
The multiplicities for different particle species, such as nucleons, pions, $\Lambda$-baryons and light nuclei are calculated in a box of (10 fm)$^3$ at $T=150$ MeV for 4000 events.
The box is initially filled with each 60 nucleons and pions.

One can observe in Figure \ref{fig:boxtest} the formation of light nuclei over time and find that the analytical equilibrium multiplicities are correctly reproduced by the transport approach.
We observe however a slightly slower equilibration.
Hence, we investigate technical parameters such as test particles and time step size.
Finer time steps did not change our results.
The same goes for the number of test particles where we show a calculation with double the number of test particles as denoted by the full line in Figure \ref{fig:boxtest}.
Since no difference can be observed, calculations for the following sections are performed with 10 test particles.

%#######################################################
\section{Light nuclei production \label{sec:results}}
%#######################################################

In this section we present our results for the light nuclei production in gold-gold collisions at RHIC-BES-energies.
More precisely, we investigate the production of deuterons, $^3\mathrm{He}$ nuclei, tritons and hypertritons at 7.7 GeV, 14.5 Gev and 19.6 GeV in the 10\% most central events.
We start with results for the transverse momentum spectra for each nucleus at the three different energies.
Next we present some particle ratios.
Then we have a look at the nuclei multiplicity as a function of time to see when they are formed and how fast equilibrium is reached.
To understand how light nuclei are formed, we further investigate the production mechanisms and when light nuclei production takes place.

\subsection{Transverse momentum spectra}
\label{sec:transverse_momentum}

The light nuclei transverse momentum spectra are calculated at mid-rapidity for the three different energies.
Here we compare the dynamic light nuclei formation via stochastic rates and the production via coalescence.
In both scenarios the evolution within the transport model starts based on particles sampled on the Cooper-Frye hypersurface of a hydrodynamic evolution.
For the dynamic production the sampled particles include light nuclei.
They undergo interactions and can be destroyed or newly formed in this stage.

However in the coalescence scenario no light nuclei are sampled on the Cooper-Frye hypersurface.
In the afterburner stage rescattering also takes place but the nuclei are not present.
Instead they are formed in the final sate of the afterburner out of nucleons based on their distance on phase-space (see Section \ref{sec_coalescence}).

For comparison, we additionally calculate the spectra of nuclei sampled on the hydrodynamic hypersurface without hadronic rescattering in the afterburner stage.
The impact of the afterburner stage is illustrated this way.

\begin{figure}
	\centering
	\begin{subfigure}{\columnwidth}
		\includegraphics[width=\textwidth]{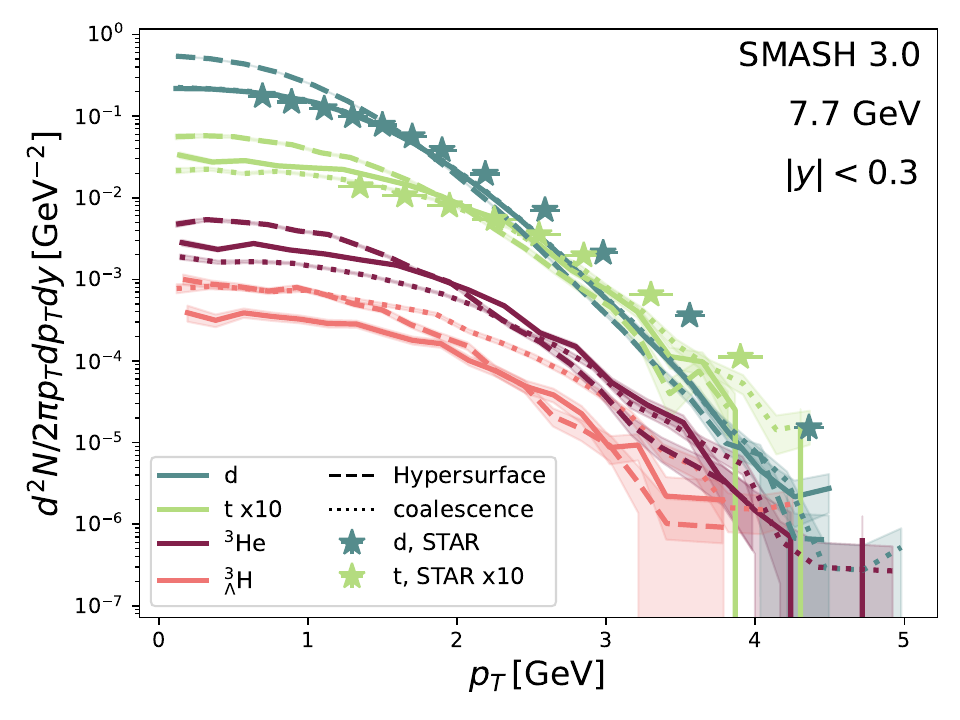}
	\end{subfigure}
	\begin{subfigure}{\columnwidth}
		\includegraphics[width=\textwidth]{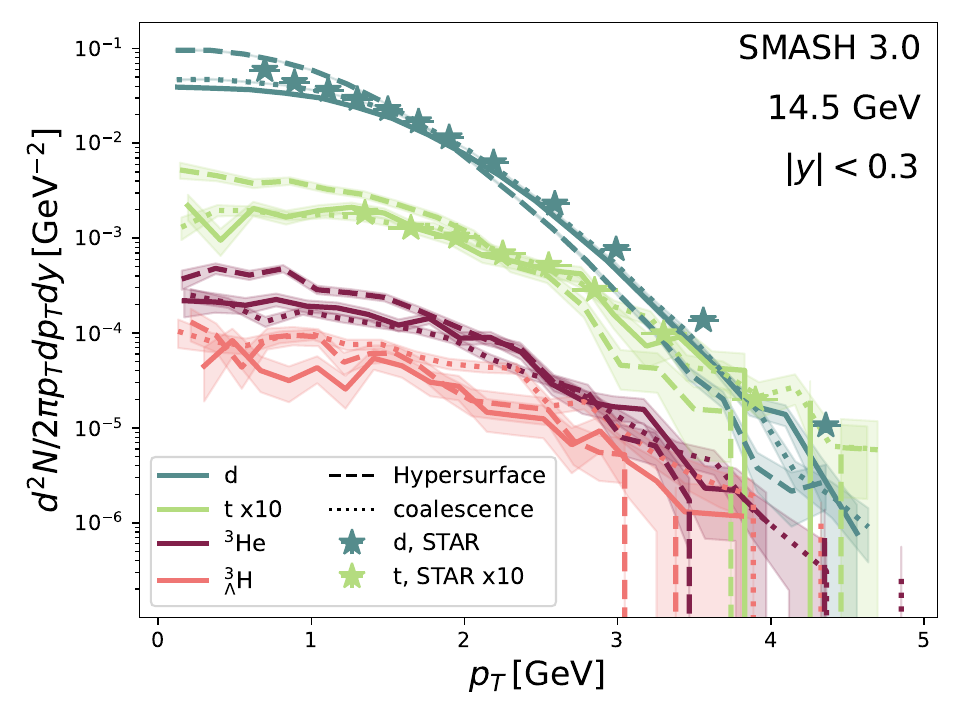}
	\end{subfigure}
	\begin{subfigure}{\columnwidth}
		\includegraphics[width=\textwidth]{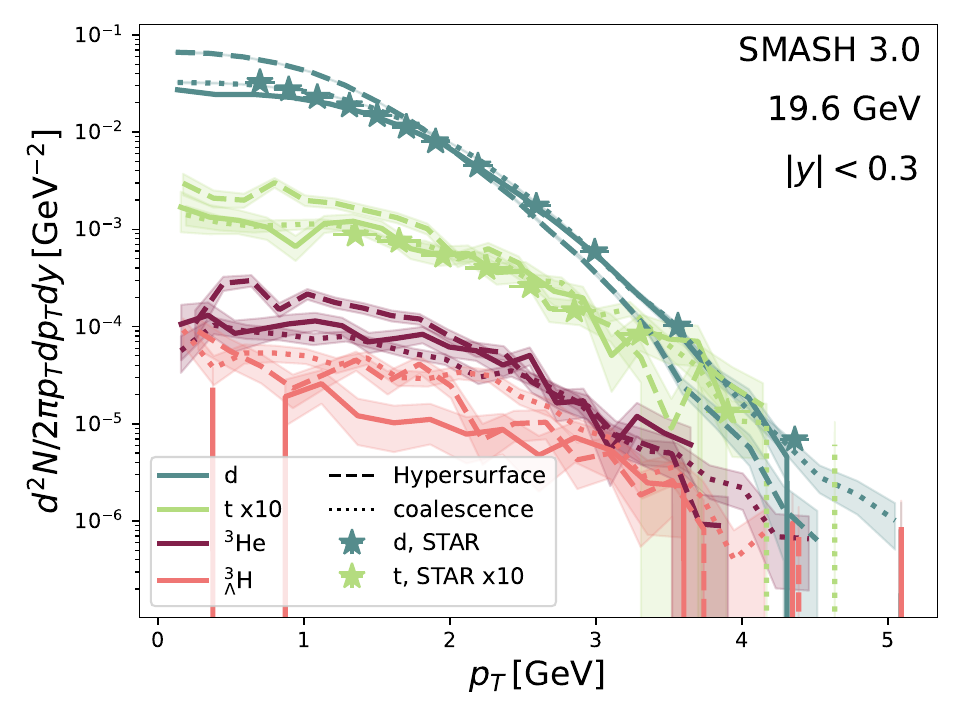}
	\end{subfigure}
	\caption{Transverse momentum spectra for different light nuclei species at mid-rapidity at different energies in central Au-Au collisions. Data points are taken from \cite{STAR:2019sjh} and \cite{STAR:2022hbp}.}
    \label{fig:pt}
\end{figure}

The transverse momentum spectra for the three different energies are compared to data from the STAR collaboration in figure \ref{fig:pt}. 
One can find, that both, coalescence and dynamic production describe the data points well for all energies and each particle species with a slight undershoot for higher $p_{T}$ at lower collision energy.
The slopes of the nuclei spectra without any rescattering (dashed curves) are too steep and the data clearly favor the calculations including the hadronic afterburner stage.
We conclude that this stage is important to obtain realistic light nuclei spectra as it is the case for proton spectra (see Petersen:2014yqa and references therein).

\subsection{Particle ratios\label{sec:ratios}}

The STAR collaboration has published ratios of protons and light nuclei multiplicities such as $N_d/N_p$, $N_t/N_p$ and $N_{t}N_{p}/N_{d}^{2}$ \cite{STAR:2022hbp}.
Calculations with the coalescence model show that the ratios are sensitive to fluctuations of the baryon density, which in turn appear close to the critical end point or a first order phase transition \cite{Sun:2017xrx,Sun:2018jhg}.
The ratios can further be used to extract information on the interaction of nucleons \cite{Shuryak:2018lgd,Shuryak:2019ikv}.

\begin{figure}
    \centering
    \includegraphics[width=\columnwidth]{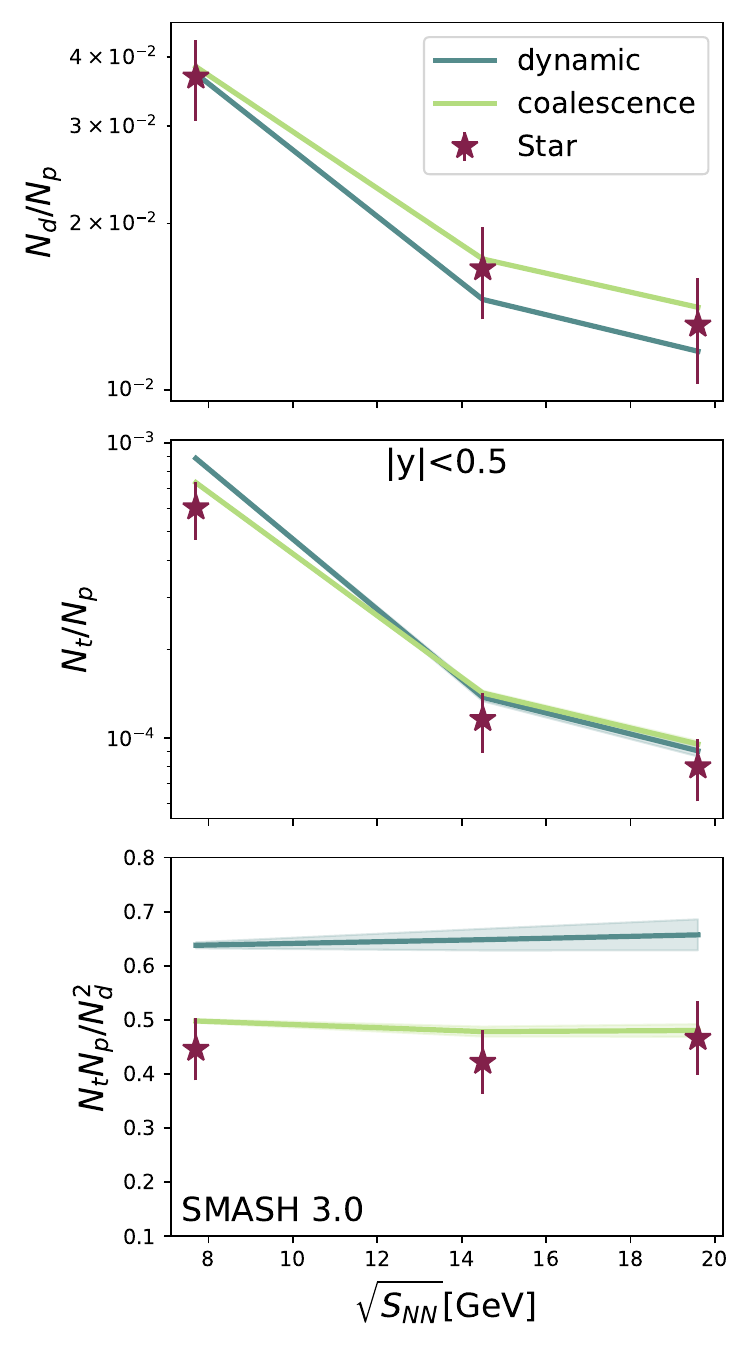}
    \caption{Collision energy dependence of particle ratios at mid-rapidity in central Au-Au collisions. Data points are taken from \cite{STAR:2022hbp}.}
    \label{fig:ratios}
\end{figure}

In Figure \ref{fig:ratios} one can find three different particle ratios.
These ratios are calculated at the end of the afterburner at mid-rapidity for different energies.
The two upper plots show the single ratios for the number of deuterons and tritons divided by the number of protons respectively.
In the lower plot the double ratio $N_{t}N_{p}/N_{d}^{2}$ is presented.

Comparing our calculations to STAR data \cite{STAR:2022hbp} we can see, that the single ratios are well described by both the dynamic production and the coalescence model.
The motivation to investigate the double ratio is that it can be related to critical fluctuations.
For both calculations we observe an energy independent double ratio. 
The coalescence model performs better for the double ratio but we would like to emphasize, that critical fluctuations are not implemented in the model.

\subsection{Time dependence of light nuclei multiplicities\label{sec:multiplicities}}

In this section we present the light nuclei multiplicities as a function of time at 7.7, 14.5 and 19.6 GeV.
Multiplicities for deuterons, tritons and hypertritons are shown for mid-rapidity and the full phase space.
Calculations for $^3\mathrm{He}$ nuclei yield very similar results as tritons and are hence presented in Appendix \ref{ap_he3}.
For each particle and each energy we compare two curves: the solid curve represents the nuclei produced in multi-particle reactions in the afterburner stage.
The dashed curve denotes the nuclei, sampled on the hydrodynamic Cooper-Frye hypersurface without performing the afterburner stage.

\begin{figure}
	\centering
	\begin{subfigure}{\columnwidth}
		\includegraphics[width=\textwidth]{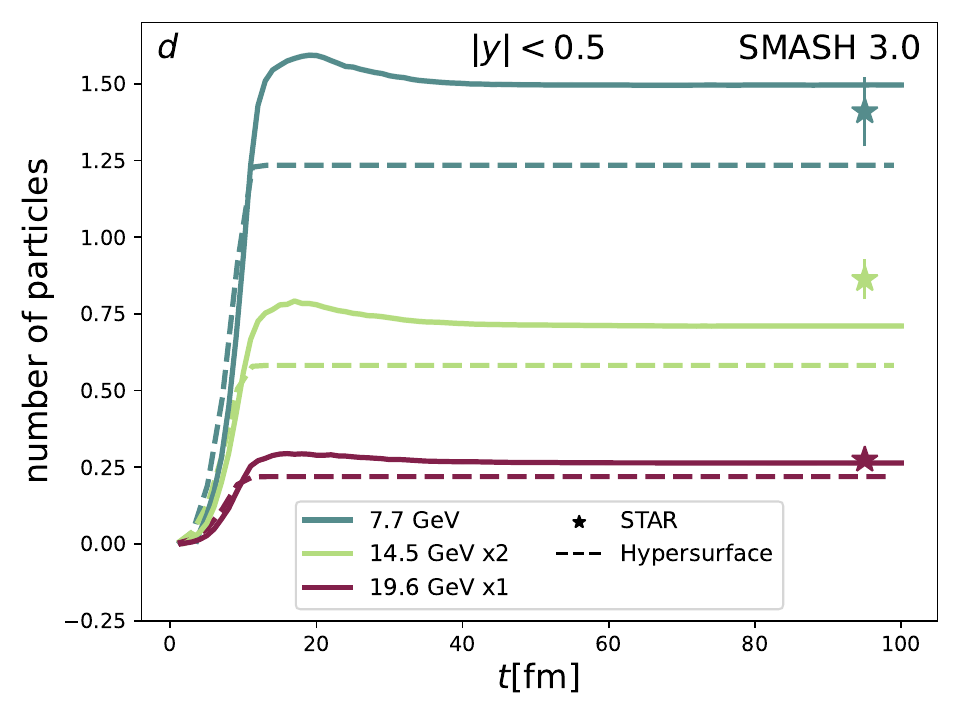}
	\end{subfigure}
	\begin{subfigure}{\columnwidth}
		\includegraphics[width=\textwidth]{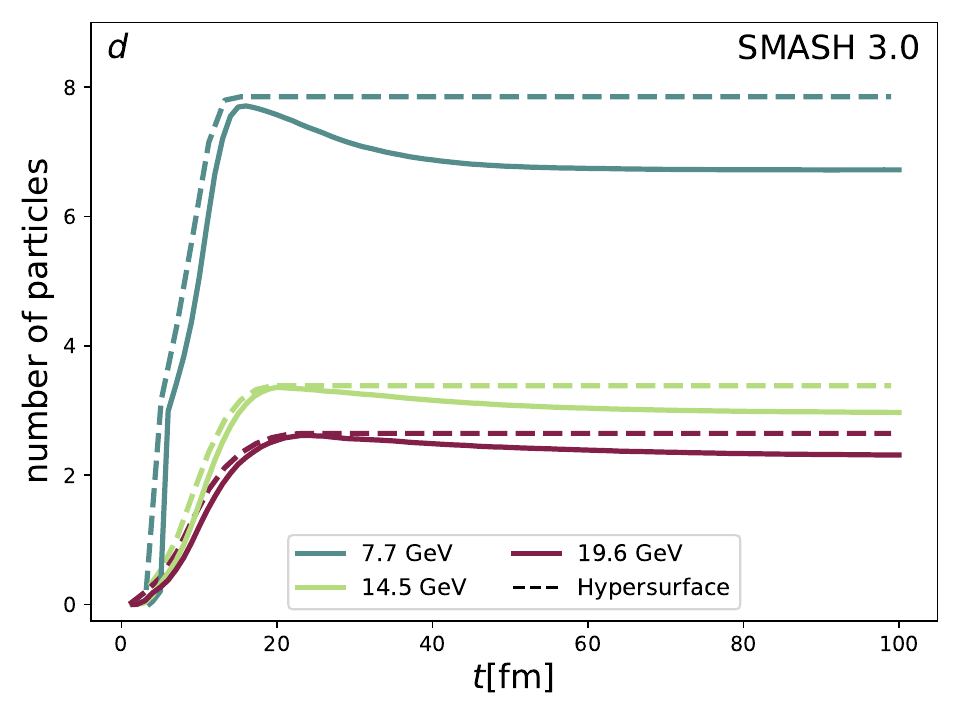}
	\end{subfigure}
	\caption{Multiplicities of deuterons at different energies in central Au-Au collisions as a function of time.
    The full lines represent the full afterburner calculations and the dashed lines are the multiplicities sampled on the Cooper-Frye hypersurface of the hydrodynamic calculation.
    Data points are taken from \cite{STAR:2019sjh}.}
    \label{fig:mult_d}
\end{figure}

Figure \ref{fig:mult_d} shows the deuteron yield as a function of time together with the experimentally observed yield in the final state \cite{STAR:2019sjh}.
At mid-rapidity one can observe for all energies that the dynamic production of deuterons leads to a higher number of nuclei compared to the number given of sampled particles on the hydrodynamic hypersurface.
This means, that at mid-rapidity, more nuclei are produced than destroyed in the afterburner stage.
This could be caused by a large number of protons and neutrons at mid-rapidity so that the deuteron production can take place frequently.
The agreement with experimental data is improved by catalysis reactions in the afterburner stage.
In the $4\pi$ scenario the situation changes.
The afterburner multiplicity stays slightly below the number of particles from the hypersurface in the beginning of the evolution and the amount of light nuclei is reduced at later times.
This can be understood by taking into account that the density decreases, as the system expands.
The destruction processes outweigh the formation processes as a consequence.
Section \ref{sec:productionmech} contains a detailed analysis of the corresponding production rates.
In summary, the total number of deuterons is reduced as the system expands but more deuterons are formed at mid-rapidity.

\begin{figure}
	\centering
	\begin{subfigure}{\columnwidth}
		\includegraphics[width=\textwidth]{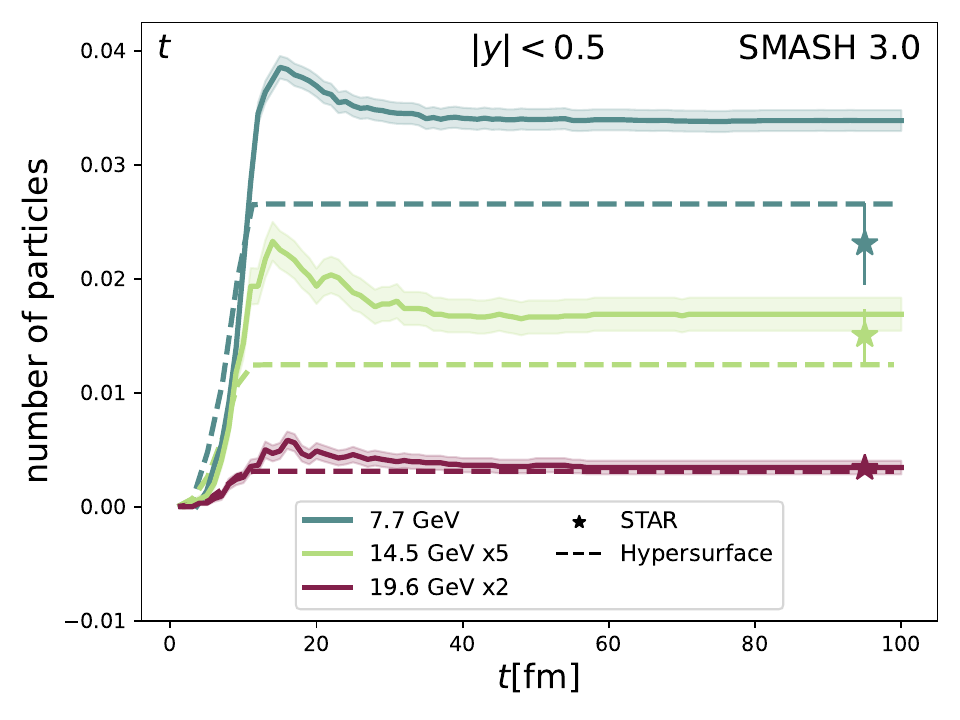}
	\end{subfigure}
	\begin{subfigure}{\columnwidth}
		\includegraphics[width=\textwidth]{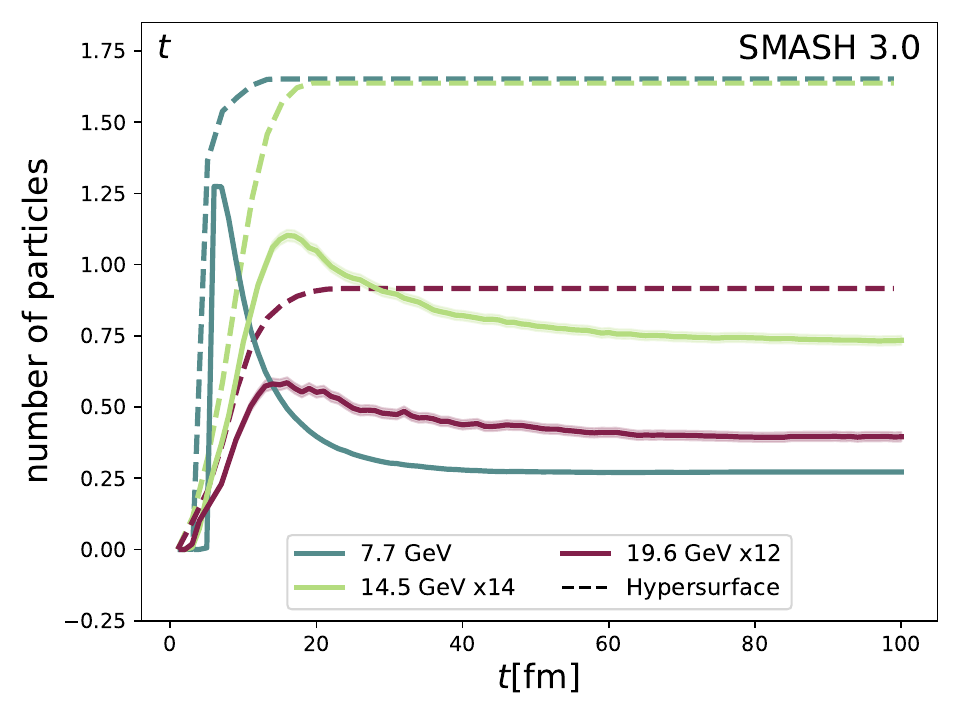}
	\end{subfigure}
	\caption{Multiplicities of tritons at different energies in central Au-Au collisions as a function of time.
    The full lines represent the full afterburner calculations and the dashed lines are the multiplicities sampled on the Cooper-Frye hypersurface of the hydrodynamic calculation.
    Data points are taken from \cite{STAR:2022hbp}.}
    \label{fig:mult_t}
\end{figure}

We continue with the triton multiplicities as a function of time as presented in Figure \ref{fig:mult_t}.  
At mid-rapidity, we can observe a similar behaviour as for the deuteron.
The increase at early times is in this case slightly more pronounced and the reduction of the mid-rapidity multiplicity towards later times is larger.
The data points are well reproduced for the higher energies but the afterburner calculation overshoots the measurement at 7.7 GeV as the afterburner increases the mid-rapidity yield at this energy, where the net-baryon density is the largest.
At $4\pi$ the afterburner reduces the yields drastically, even stronger for lower energies.
We would like to highlight that for this energy at 4pi the deviation from the hydro curve is immense compared to the mid-rapidity region.
The behavior for the $^3\mathrm{He}$ nuclei is very similar to the triton so that the results are presented in Appendix \ref{ap_he3}.

In figure \ref{fig:mult_hy} one can find the multiplicities for the hypertritons.
At mid-rapidity the afterburner curves exceed the hydrodynamic curves at earlier times like it was the case for the other nuclei.
Considering the statistical uncertainties, the afterburner calculations reach for all energies the multiplicities from the hypersurface at later times.
Overall, we observe at mid-rapidity the same behaviour as for d, t and $^3\mathrm{He}$ except the reduction of the hypertriton yield brings the multiplicity down to the number of particles from the hydrodynamic hypersurface whereas the yield of other nuclei increases due to the afterburner stage.
Hypertritons at $4\pi$ show a unique feature at 7.7 GeV: For other light nuclei, the multiplicity is reduced during the afterburner stage.
However, for the $_\Lambda^3H$ more nuclei are produced in the beginning and then the multiplicity converges to the number of particles on the hydrodynamic hypersurface at later times.

To summarize these results, lets record that at $4\pi$ the afterburner reduces the amount of nuclei at later times for all energies and all nuclei species, except the 7.7 GeV hypertriton curve.
This reduction is eminently pronounced for triton and $^3\mathrm{He}$ nuclei.
At mid-rapidity the dynamical production leads to a higher amount of nuclei in the final state except for the hypertriton.
In this case the multiplicities equilibrate to the hydrodynamic curves.

A possible interpretation of the unique behaviour of the hypertriton may be the following.
In general, $4\rightarrow 2$ reactions become less likely as the system expands compared to $2\rightarrow4$ reactions.
This naturally leads to a reduction of the $A=3$ nuclei yield. 
The hypertriton is mainly formed in the reaction $\Lambda pn\pi\rightarrow^3_\Lambda\mathrm{H}\pi$, so that a $\Lambda$ is required for the production.
Compared to nucleons, the $\Lambda$ has only small hadronic cross sections, so that the probability to participate in a process to form a hypertriton is increased.
This may lead to the observed enhancement of the hypertriton yield.
A further aspect to consider is how many nucleons and $\Lambda$ baryons are present in an excited state.
Resonances do not participate in reactions to form or destroy light nuclei.

%deuterons

%tritons

%hypertriton
\begin{figure}
	\centering
	\begin{subfigure}{\columnwidth}
		\includegraphics[width=\textwidth]{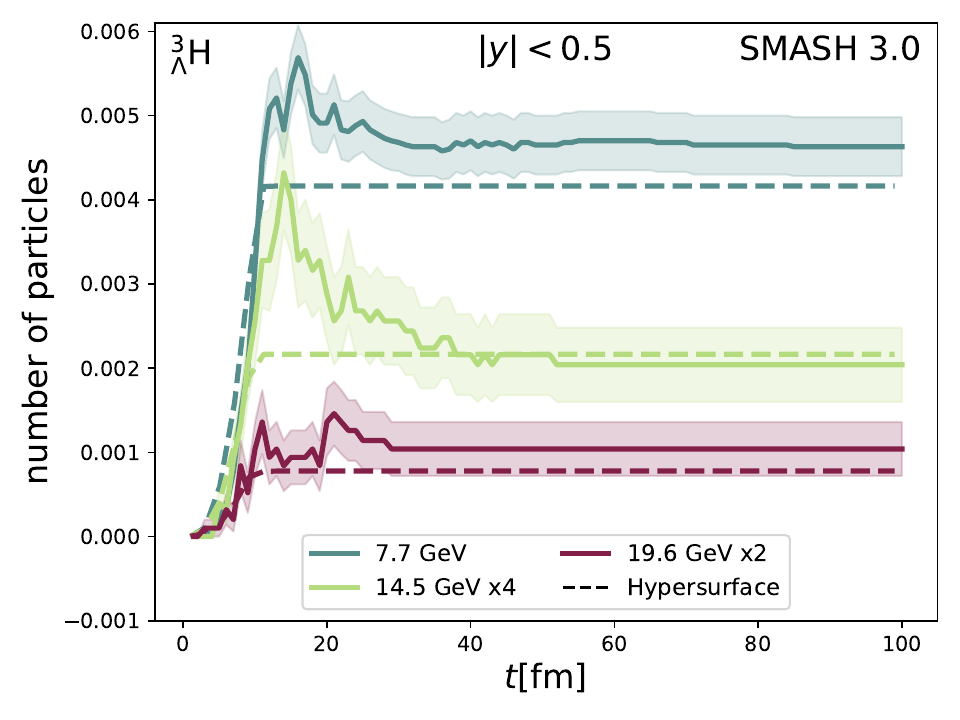}
	\end{subfigure}
	\begin{subfigure}{\columnwidth}
		\includegraphics[width=\textwidth]{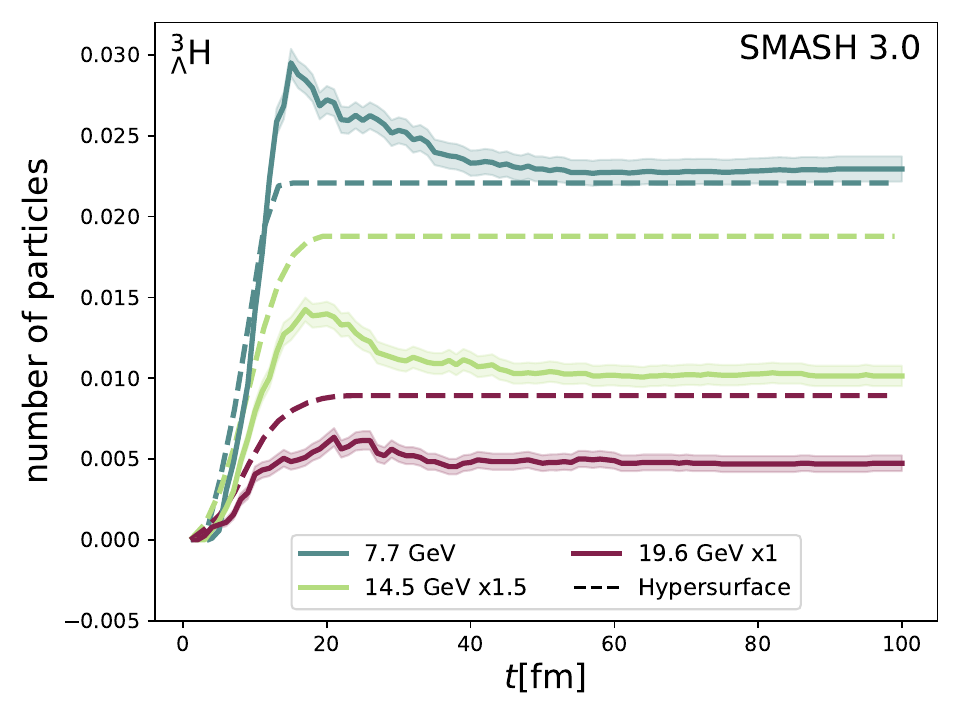}
	\end{subfigure}
	\caption{Multiplicities of hypertritons at different energies in central Au-Au collisions as a function of time.
    The full lines represent the full afterburner calculations and the dashed lines are the multiplicities sampled on the Cooper-Frye hypersurface of the hydrodynamic calculation.}
    \label{fig:mult_hy}
\end{figure}

\begin{figure}
    \centering
    \includegraphics[width=\columnwidth]{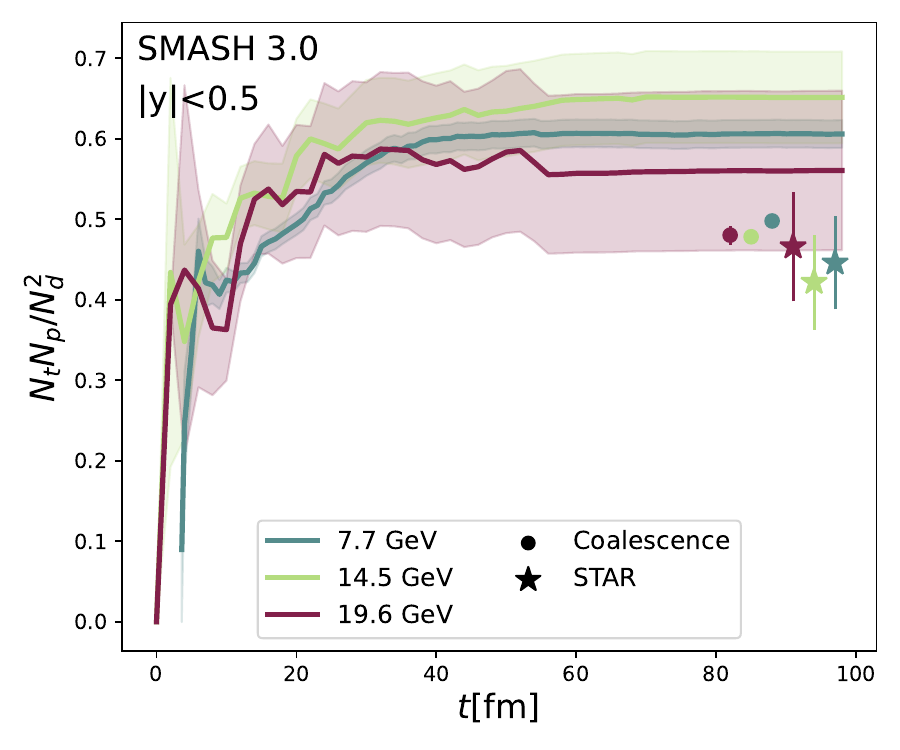}
    \caption{The double ratio $N_{t}N_{p}/N_{d}^{2}$ in Au-Au collisions at mid-rapidity is calculated as a function of time. Three different energies are shown. The curves are compared to the the ratios obtained by coalescence at the end of the afterburner (circles) and to data from the STAR collaboration\cite{STAR:2022hbp} (stars).}
    \label{fig:ratio3}
\end{figure}

Based on the results presented in this section, we can calculate the time dependence of the particle ratios discussed in Section \ref{sec:ratios}.
In Figure \ref{fig:ratio3} the time evolution of the double ratio $N_{t}N_{p}/N_{d}^{2}$ is shown.
At all energies the ratios behave in a similar way: the curves rise in the beginning and saturate towards later times.
The similarity between the curves is not surprising, considering the energy independence observed in the lower panel of Figure \ref{fig:ratios}.
The non-trivial time dependence of deuteron and triton yields observed in Figures \ref{fig:mult_d} and \ref{fig:mult_t} cancel out in the double ratio, as the curves rise monotonically.
Figure \ref{fig:ratio3} also includes the experimentally measured ratios\cite{STAR:2022hbp} and the once obtained from final state coalescence.
Like in Figure \ref{fig:ratios} the coalescence and the experimental data points coincide while the dynamic production overshoots slightly.

\subsection{Production mechanisms\label{sec:productionmech}}
\begin{figure}
    \centering
    \includegraphics[width=\columnwidth]{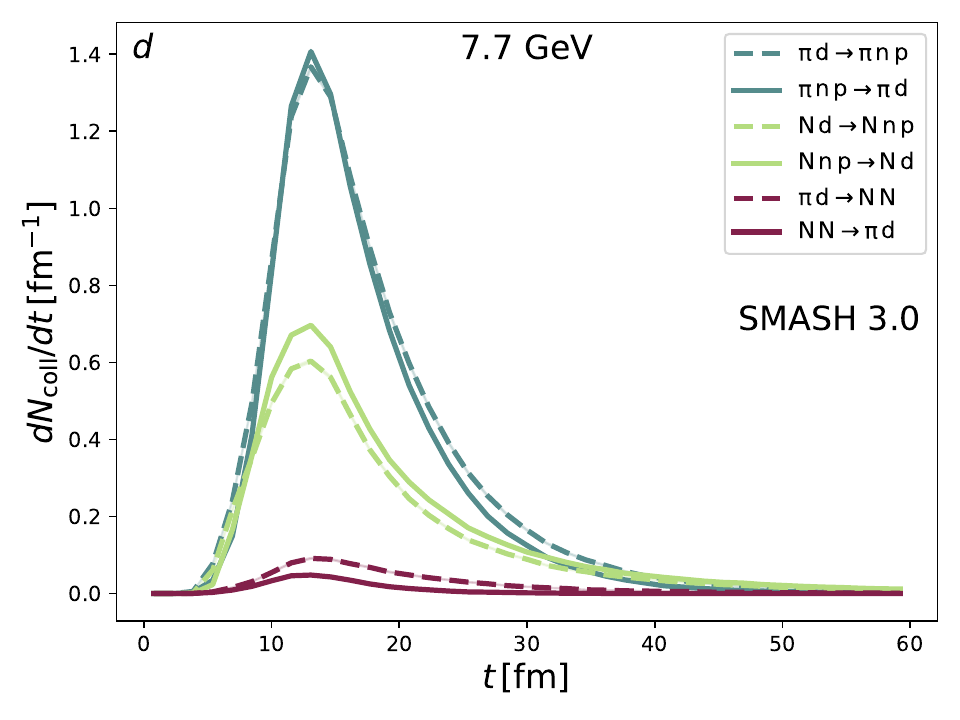}
    \caption{Scattering rates for deuterons at 7.7 GeV in central Au-Au-collisions. The solid curves represent formation mechanisms while the dashed curves represent disintegration mechanisms.}
    \label{fig:coll_d_7}
\end{figure}

\begin{figure}
    \centering
    \includegraphics[width=\columnwidth]{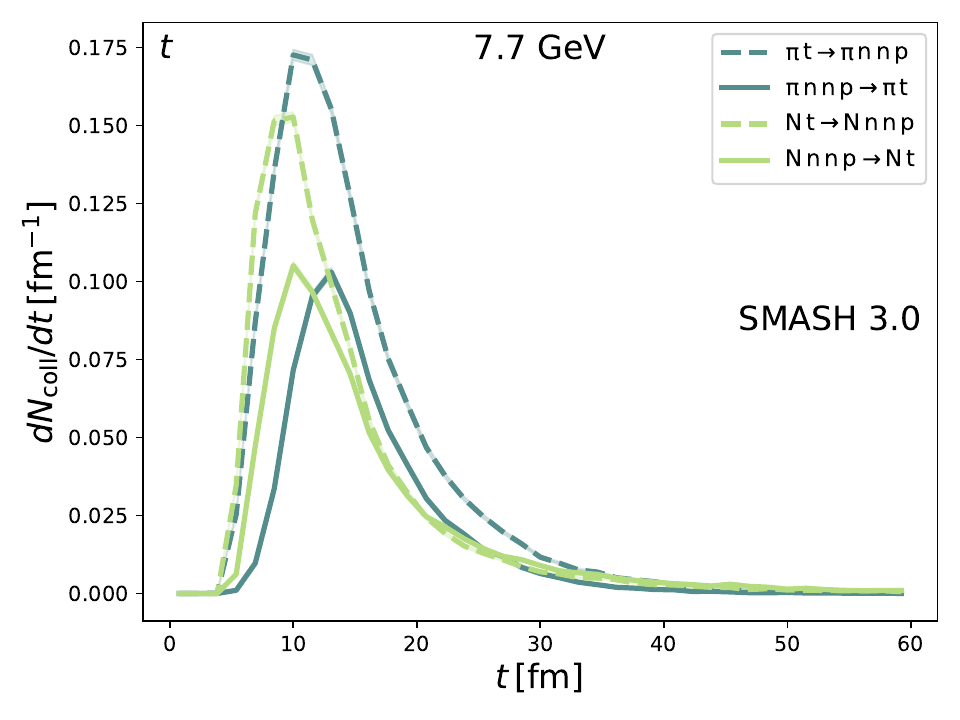}
    \caption{Scattering rates for tritons at 7.7 GeV in central Au-Au-collisions. The solid curves represent formation mechanisms while the dashed curves represent disintegration mechanisms.}
    \label{fig:coll_t_7}
\end{figure}

\begin{figure}
    \centering
    \includegraphics[width=\columnwidth]{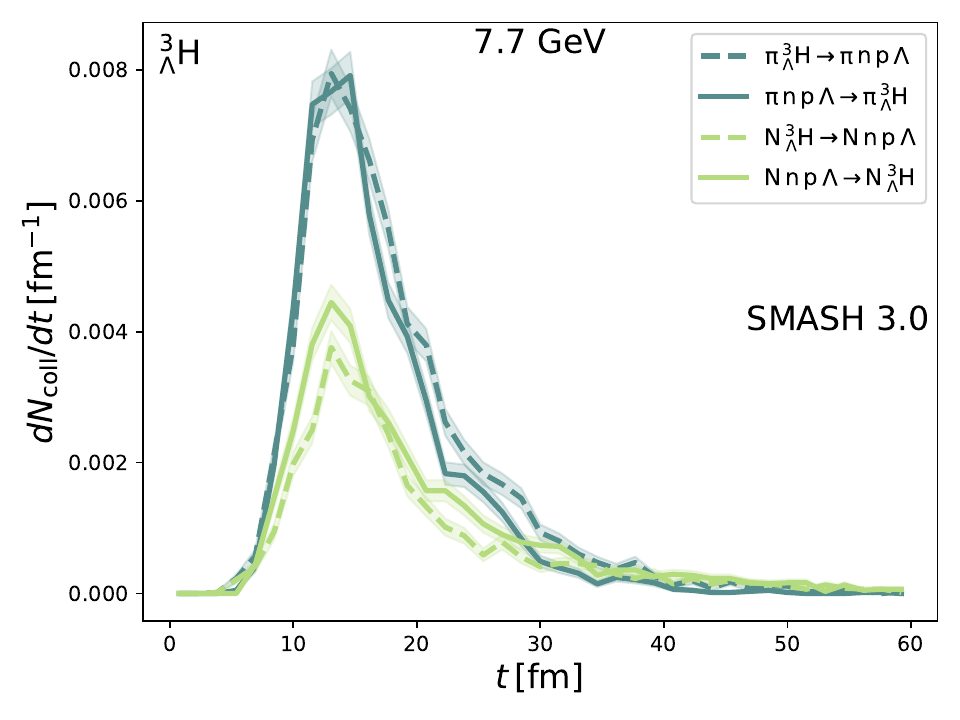}
    \caption{Scattering rates for hypertritons at 7.7 GeV in central Au-Au-collisions. The solid curves represent formation mechanisms while the dashed curves represent disintegration mechanisms.}
    \label{fig:coll_hy_7}
\end{figure}

We will now have a look on the light nuclei production mechanisms to get a better understanding about the formation process during the afterburner.
This is motivated by the fact, that we cannot retrace the reactions in the afterburner properly by only knowing the yields. 
For example a constant number of particles can either be achieved by no reactions or by as many forward as backward reactions. 
Also a small reduction of the nuclei yield during the afterburner, as the case for the deuterons in $4\pi$ (see figure \ref{fig:mult_d}), does not imply that there are only few reactions. 
To illustrate the production mechanisms we calculate the collision rates of the different reactions for the formation and disintegration of deuterons, tritons and hypertritons. 
The rates are shown in Figures \ref{fig:coll_d_7}, \ref{fig:coll_t_7} and \ref{fig:coll_hy_7} for the lowest presented energy 7.7 GeV.
Here most of the hadronic rescattering takes place in the first
30 to 40 fm and the maximum is located around 10 to 15 fm for each particle.

For deuterons (see Figure \ref{fig:coll_d_7}) one can learn that in the first 20 fm the formation and disintegration processes are roughly balanced.
At later times the disintegration outweighs slightly.
This is consistent with the 7.7 GeV curve in Figure \ref{fig:mult_d}, where the yield of dynamically produced nuclei in the final state at $4\pi$ is smaller than the yield on the Cooper-Frye hypersurface.
The small rates of the $NN\leftrightarrow \pi d$ reactions verify that the multiparticle reations are the most important contribution for the light nuclei production.
For the production of deuterons, the pion catalysis reactions are most frequent.
In Appendix \ref{ap_higher_energies} we show rates for higher collision energies, where we observe that with increasing energy the pion catalysis reactions become more dominant.
This can be attributed to the fact that the population of pions is larger at higher energies.

The reaction for the formation and disintegration of tritons is presented in Figure \ref{fig:coll_t_7}.
The disintegration of tritons clearly dominates and therefore the rescattering reduces the yields drastically.
This is in accord with the final yield at 7.7 GeV in Figure \ref{fig:mult_t}.
For the tritons we observe that the nucleon catalysis reactions are as important as the pion catalysis reactions at this energy.
For higher energies we observe, again, that pion catalysis becomes more important (see Appendix \ref{ap_higher_energies}).

The reaction rates relevant for the hypertriton production are shown in Figure \ref{fig:coll_hy_7}.
We observe slightly more formation than disintegration in the first 20 fm which is why the afterburner stage increases the $4\pi$ yield of hypertritons (Figure \ref{fig:mult_hy}) in this time frame.
Between 20 and 40 fm the disintegration outweighs slightly.
This leads to the same multiplicity on the Cooper-Frye hypersurface as in the final state of the afterburner.
For the hypertriton, the pion catalysis outweighs the nucleon catalysis rections already at 7.7 GeV. Collision rates at higher energies can be found in Appendix \ref{ap_higher_energies}.

\begin{figure}
    \centering
    \includegraphics[width=\columnwidth]{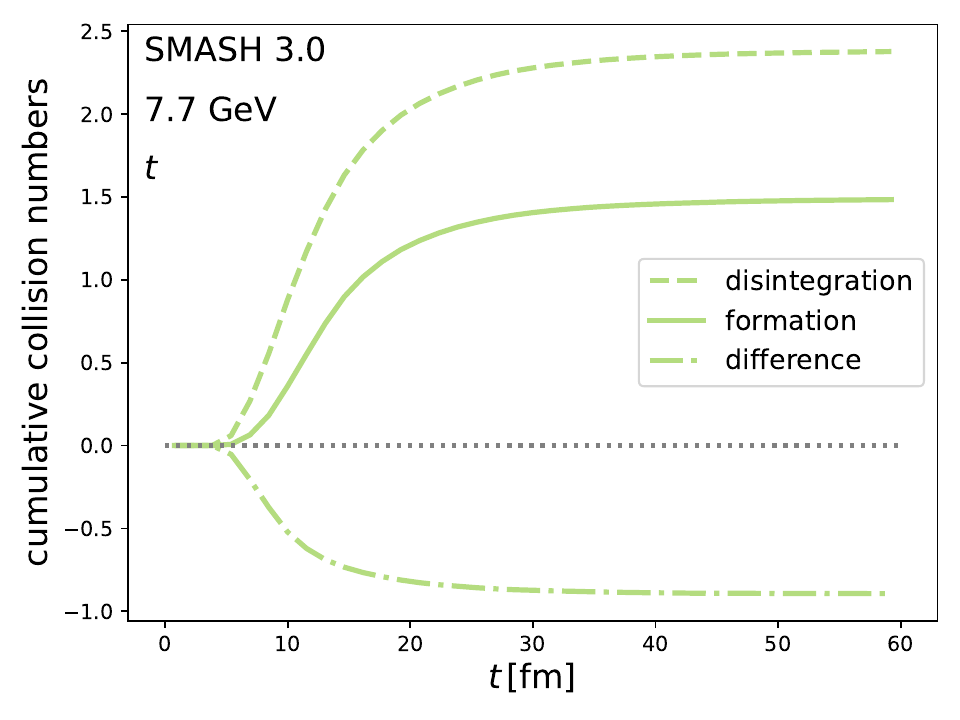}
    \caption{Cumulative collision numbers for disintegration and formation reactions for tritons at 7.7 GeV are shown. They are compared to the difference between these two curves to illustrate the change in the final particle yield.}
    \label{fig:coll_diff_t_7}
\end{figure}

\begin{figure}
    \centering
    \includegraphics[width=\columnwidth]{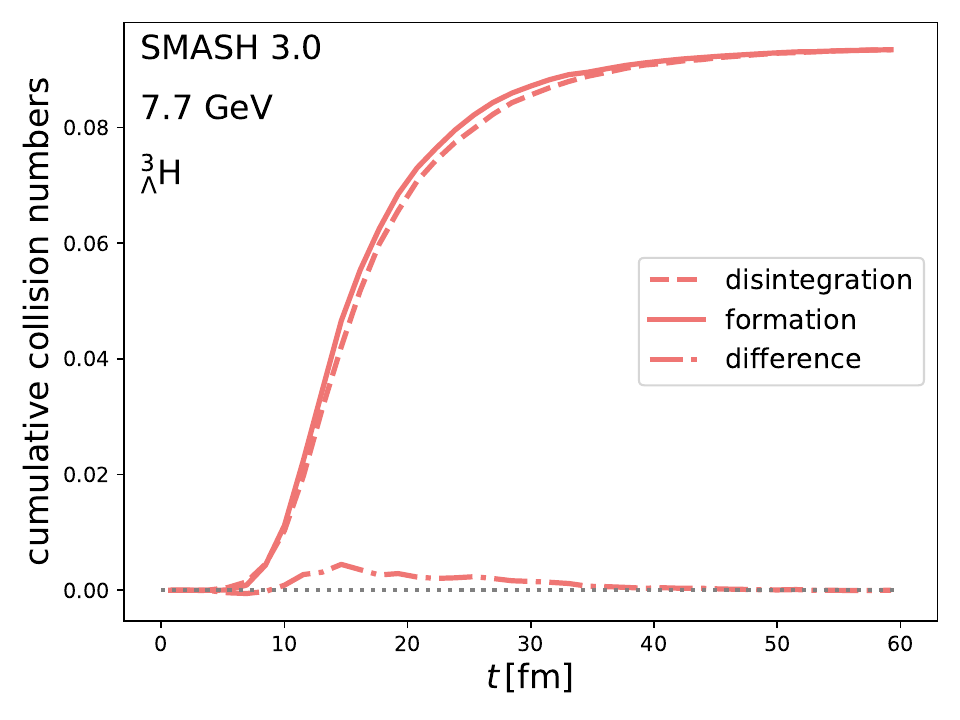}
    \caption{Cumulative collision numbers for disintegration and formation reactions for hypertritons at 7.7 GeV are shown. They are compared to the difference between these two curves to illustrate the change in the final particle yield.}
    \label{fig:coll_diff_hy_7}
\end{figure}

In Figures \ref{fig:coll_diff_d_7} to \ref{fig:coll_diff_hy_7} one can find the cumulative collision numbers for the formation processes compared to the disintegration processes to point out the change in the final particle yield.
The difference between the cumulative production number and destruction number corresponds to the change in the yield of the considered nucleus compared to the particle number sampled on the hypersurface.
We observe that the difference, and hence the impact of the afterburner on the multiplicity, is relatively small for deuterons and hypertritons.
This does not mean that the afterburner stage can be disregarded as we see that the number of reactions that involve light nuclei is large and significantly contributes to the dynamics of the system.
The importance of the afterburner stage has been demonstrated already in Section \ref{sec:transverse_momentum}.
For the triton, a reasonable number of formation reactions takes place, but the disintegration processes dominate, leading to the reduced triton yield in the final state of the afterburner stage.

\begin{figure}
    \centering
    \includegraphics[width=\columnwidth]{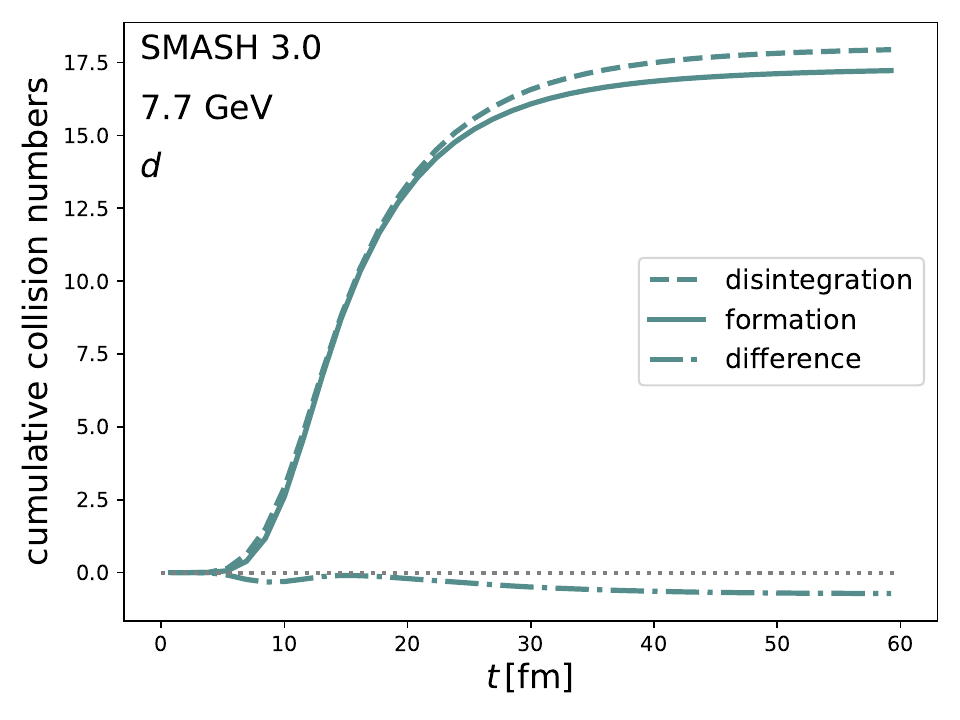}
    \caption{Cumulative collision numbers for disintegration and formation reactions for deuterons at 7.7 GeV are shown. They are compared to the difference between these two curves to illustrate the change in the final particle yield.}
    \label{fig:coll_diff_d_7}
\end{figure}

%#######################################################
\section{Summary and conclusion} \label{sec:conclusions}
%#######################################################
In this work, we have investigated the dynamic production of deuterons, tritons, $^3\mathrm{He}$ nuclei and hypertritons in the afterburner stage of a hybrid approach.
The afterburner calculation was performed using the transport model SMASH, where the nuclei are formed via multi-particle reactions realized with the stochastic collision criterion or via coalescence.

For validation, the multiplicities obtained using multi-particle reactions in a box calculation were compared to results from rate equations.
We found that the equilibrium multiplicities are correctly reproduced and a slightly slower equilibration as compared to the rate equation is observed.

In the next step, transverse momentum spectra of light nuclei are confronted with experimental data.
We found a good agreement with the measurement for the dynamic production of light nuclei and for the coalescence approach, as long as the afterburner stage is performed.

We further presented results for the particle ratios $N_d/N_p$, $N_t/N_p$ and $N_t N_p/N_d^2$.
For the single ratios, a good agreement with the experimental data is found with both the dynamic production of nuclei and the coalescence approach.
The double ratio is better described by the coalescence approach but the double ratio was proposed to be sensitive to the critical point which we do not include in the model.

To get a better understanding of the time evolution of the light nuclei production, we presented the multiplicity of light nuclei as a function of time at mid-rapidity and for the full phase-space.
Considering the full phase-space, we found that the yield of light nuclei decreases in general as the system expands, except for the hypertriton which experiences a slight increase in multiplicity at $\sqrt{s}_{NN}=7.7\,\mathrm{GeV}$ which relaxes back to the yield on the hydrodynamic Cooper-Frye hypersurface towards later times.
In contrast to the $4\pi$ multiplicities, light nuclei multiplicities tend to increase at mid-rapidity during the afterburner stage and a good agreement with the experimentally measured yields at mid-rapidity is found.

Finally, the scattering rates for the formation and disintegration of light nuclei were studied.
As in the considered systems, the number of pions is large, we observe that the pion catalysis reactions are most important for the formation of light nuclei in many cases.
For deuterons and hypertritons the relatively small change in the multiplicities does not implicit a small interaction number.
Instead, forward and backward reactions are frequent and the rates are roughly balanced.
For the tritons we observe a larger number of disintegration compared to the formation processes, which is consistent with the observed decrease in triton multiplicity in the afterburner stage.
Multi-particle reactions in the afterburner stage are successful at capturing the dynamics and production of light nuclei.

This study is focused on three energies from the RHIC Beam Energy Scan II program.
In future work, an extension to further energies is possible.
%#######################################################

\begin{acknowledgments}
The authors thank Dmytro Oliinychenko for his contribution during the initial state of this work and his helpful comments.
The authors further thank Chun Shen for providing the hydrodynamic particlization hypersurfaces.
JM and HE acknowledge the support by the State of Hesse within the Research Cluster ELEMENTS (Project ID 500/10.006).
Computational resources have been provided by the Center for Scientific Computing (CSC) at the Goethe-University of Frankfurt.
\end{acknowledgments}

%appendix
\appendix
\section{Rate equations}
\label{ap_rate_equations}
We validated the newly introduced multi-particle reactions to create $A=2$ and $A=3$ nuclei in Section \ref{sec:validation}.
For this purpose we calculated the multiplicities of different particle species as a function of time in a periodic box (see Figure \ref{fig:boxtest}).
To verify these calculations we compared them to the analytical solutions, which are obtained by solving the corresponding rate equations.
In Equation \ref{eq:implementedreactions} the implemented reactions are shown.
\begin{equation}
\begin{split}
    \pi pn\leftrightarrow\pi d\quad&\quad\pi ppn\leftrightarrow\pi\text{He}^{3}\\
    Npn\leftrightarrow Nd\quad&\quad Nppn\leftrightarrow N\text{He}^{3}\\
    \pi pnn\leftrightarrow\pi t\quad&\quad\pi pn\Lambda\leftrightarrow\pi^{3}_{\Lambda}\text{H}\\
    Npnn\leftrightarrow Nt\quad&\quad Npn\Lambda\leftrightarrow N^{3}_{\Lambda}\text{H}\\
    \label{eq:implementedreactions}
\end{split}
\end{equation}
The equation $NN\leftrightarrow\pi d$ is also implemented in the model but for the calculations in Figure \ref{fig:boxtest} it is switched off, as it is not included in the rate equations.
The rate equations read as follows
\begin{equation}
\begin{split}
    R_{d}&=n_{d}^{th}\left[\langle\sigma v_{\text{rel}}\rangle_{\pi d}n_{\pi}^{th}\lambda_{\pi}\right.\\
    &+\left.\langle\sigma v_{\text{rel}}\rangle_{pd}n_{p}^{th}\lambda_{p}+\langle\sigma v_{\text{rel}}\rangle_{nd}n_{n}^{th}\lambda_{n}\right]\\
    R_{t}&=n_{t}^{th}\left[\langle\sigma v_{\text{rel}}\rangle_{\pi t}n_{\pi}^{th}\lambda_{\pi}\right.\\
    &+\left.\langle\sigma v_{\text{rel}}\rangle_{pt}n_{p}^{th}\lambda_{p}+\langle\sigma v_{\text{rel}}\rangle_{nt}n_{n}^{th}\lambda_{n}\right]\\
    R_{^3\mathrm{He}}&=n_{^3\mathrm{He}}^{th}\left[\langle\sigma v_{\text{rel}}\rangle_{\pi ^3\mathrm{He}}n_{\pi}^{th}\lambda_{\pi}\right.\\
    &+\left.\langle\sigma v_{\text{rel}}\rangle_{p^3\mathrm{He}}n_{p}^{th}\lambda_{p}\right.+\left.\langle\sigma v_{\text{rel}}\rangle_{n^3\mathrm{He}}n_{n}^{th}\lambda_{n}\right]\\
    R_{^3_{\Lambda}\mathrm{H}}&=n_{^3_{\Lambda}\mathrm{H}}^{th}\left[\langle\sigma v_{\text{rel}}\rangle_{\pi ^3_{\Lambda}\mathrm{H}}n_{\pi}^{th}\lambda_{\pi}\right.\\
    &+\left.\langle\sigma v_{\text{rel}}\rangle_{p^3_{\Lambda}\mathrm{H}}n_{p}^{th}\lambda_{p}\right.+\left.\langle\sigma v_{\text{rel}}\rangle_{n^3_{\Lambda}\mathrm{H}}n_{n}^{th}\lambda_{n}\right]\\
    \label{eq:re1}
\end{split}
\end{equation}

\begin{equation}
\begin{split}
    n_{d}^{th}\dot{\lambda_{d}}&=R_{d}(\lambda_{p}\lambda_{n}-\lambda_{d})\\
    n_{t}^{th}\dot{\lambda_{t}}&=R_{t}(\lambda_{p}\lambda_{n}\lambda_{n}-\lambda_{t})\\
    n_{^3\mathrm{He}}^{th}\dot{\lambda_{^3\mathrm{He}}}&=R_{^3\mathrm{He}}(\lambda_{p}\lambda_{p}\lambda_{n}-\lambda_{^3\mathrm{He}})\\
    n_{^3_{\Lambda}\mathrm{H}}^{th}\dot{\lambda_{^3_{\Lambda}\mathrm{H}}}&=R_{^3_{\Lambda}\mathrm{H}}(\lambda_{p}\lambda_{n}\lambda_{\Lambda}-\lambda_{^3_{\Lambda}\mathrm{H}})\\
    n_{p}^{th}\dot{\lambda_{p}}&=-n_{d}^{th}\dot{\lambda_{d}}-n_{t}^{th}\dot{\lambda_{t}}-2n_{^3\mathrm{He}}^{th}\dot{\lambda_{^3\mathrm{He}}}\\
    &-n_{^3_{\Lambda}\mathrm{H}}^{th}\dot{\lambda_{^3_{\Lambda}\mathrm{H}}}\\
    n_{n}^{th}\dot{\lambda_{n}}&=-n_{d}^{th}\dot{\lambda_{d}}-2n_{t}^{th}\dot{\lambda_{t}}-n_{^3\mathrm{He}}^{th}\dot{\lambda_{^3\mathrm{He}}}\\
    &-n_{^3_{\Lambda}\mathrm{H}}^{th}\dot{\lambda_{^3_{\Lambda}\mathrm{H}}}\\
    n_{\Lambda}^{th}\dot{\lambda_{\Lambda}}&=-n_{^3_{\Lambda}\mathrm{H}}^{th}\dot{\lambda_{^3_{\Lambda}\mathrm{H}}}\\
    \dot{\lambda_{\pi}}&=0\,.\\
    \label{eq:re2}
\end{split}
\end{equation}

\clearpage
%multiplicities and coll rates of He3
\section{Results for $^3\rm He$ nuclei \label{ap_he3}}
\begin{figure}
    \centering
    \includegraphics[width=\columnwidth]{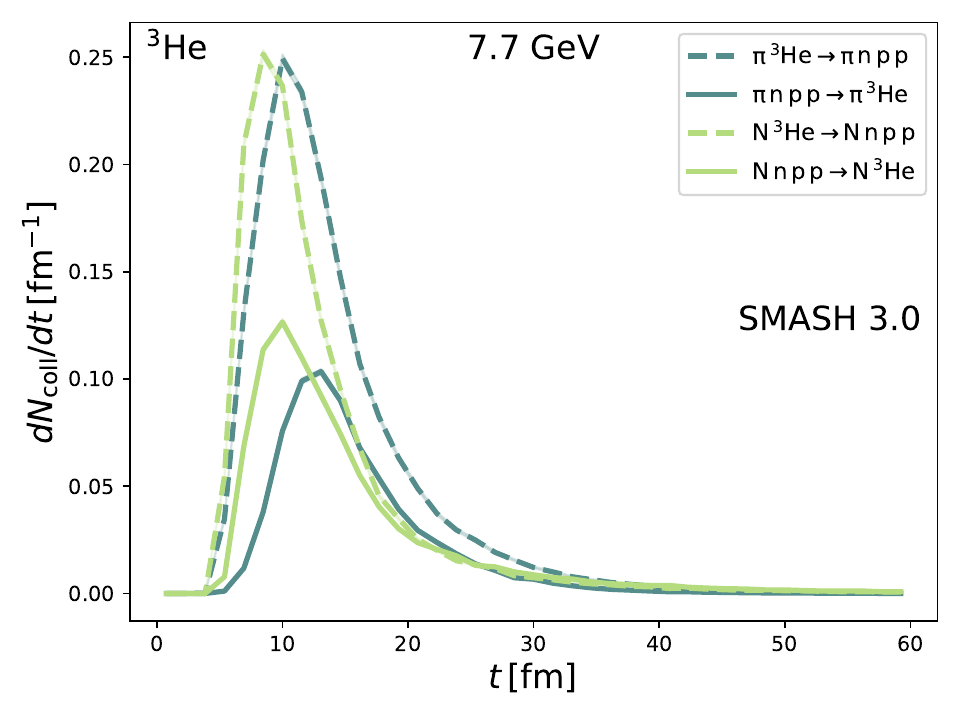}
    \caption{Scattering rates for $^3\mathrm{He}$ nuclei at 7.7 GeV in central Au-Au-collisions. The solid curves represent formation mechanisms while the dashed curves represent disintegration mechanisms.}
    \label{fig:coll_h_7}
\end{figure}

Here we present the multiplicities and the scattering rates as functions of time for $^3\mathrm{He}$ nuclei.
As the plots are looking very similar to the ones obtained for the tritons, we decided to not show them with the others in Sections \ref{sec:multiplicities} and \ref{sec:productionmech} but put them here in the appendix instead.
Figures \ref{fig:mult_h} and \ref{fig:mult_h_mid} show the $^3\mathrm{He}$ yield as a function of time at full phase space and at mid-rapidity respectively for three different energies.
For each energy two curves are compared, which are the solid curve, representing the nuclei produced in multi-particle reactions
in the afterburner stage and the dashed curve, denoting the nuclei, sampled on the hydrodynamic Cooper-Frye hypersurface without performing the afterburner stage.
\begin{figure}
    \centering
    \includegraphics[width=\columnwidth]{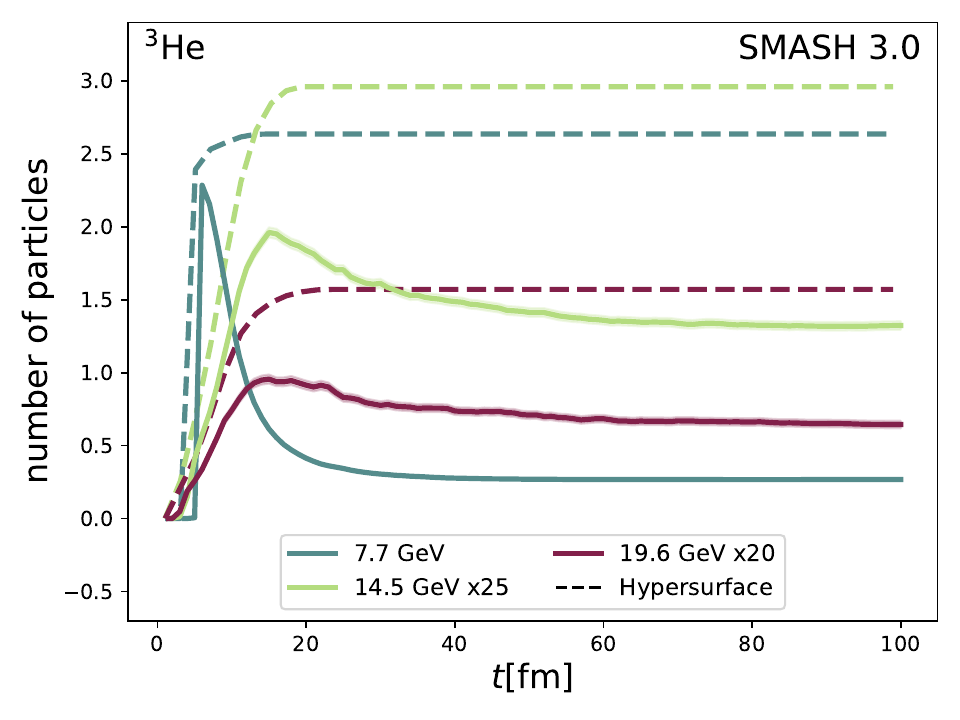}
    \caption{$4\pi$ multiplicities of $^3\mathrm{He}$ nuclei at different energies in central Au-Au collisions as a function of time.
    The full lines represent the full afterburner calculations and the dashed lines are the multiplicities sampled on the Cooper-Frye hypersurface of the hydrodynamic calculation.}
    \label{fig:mult_h}
\end{figure}
\begin{figure}
    \centering
    \includegraphics[width=\columnwidth]{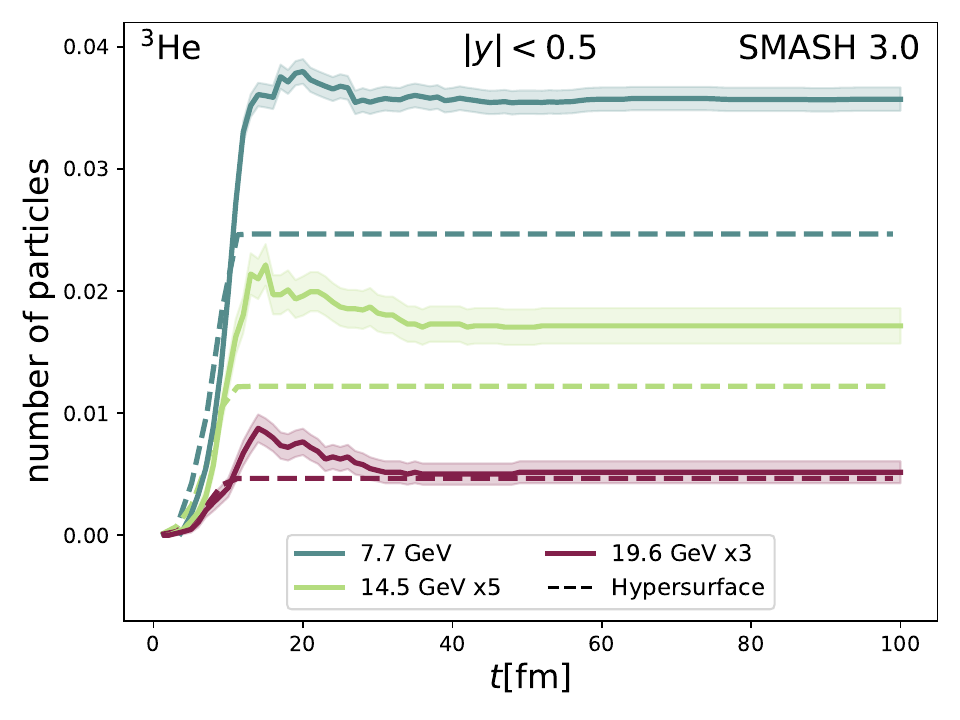}
    \caption{Mid-rapidity multiplicities of $^3\mathrm{He}$ nuclei at different energies in central Au-Au collisions as a function of time.
    The full lines represent the full afterburner calculations and the dashed lines are the multiplicities sampled on the Cooper-Frye hypersurface of the hydrodynamic calculation.}
    \label{fig:mult_h_mid}
\end{figure}
Figure \ref{fig:coll_h_7} shows the scattering rates for $^3\mathrm{He}$ nuclei at 7.7 GeV.
Nuclei formation via pion (blue curve) or nucleon (green curve) catalysis reactions is compared to disintegration and the same behavior is found as for the tritons in Section \ref{sec:productionmech}.

\clearpage
%collision rates
\section{Scattering rates at higher energies\label{ap_higher_energies}}
\begin{figure}
    \centering
    \includegraphics[width=\columnwidth]{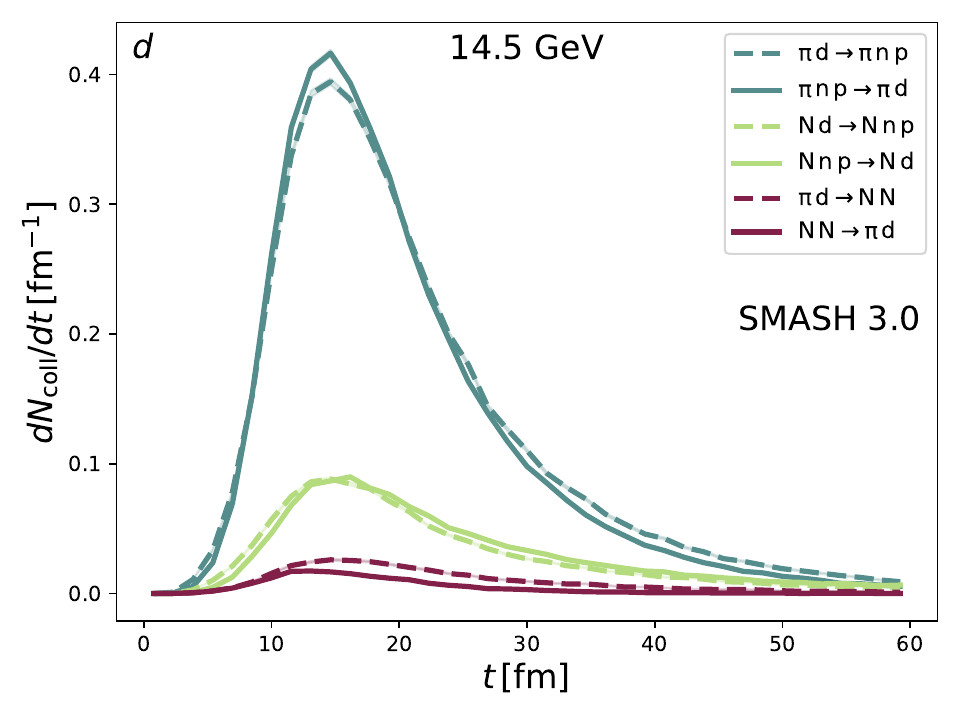}
    \caption{Scattering rates for deuterons at 14.5 GeV in central Au-Au-collisions. The solid curves represent formation mechanisms while the dashed curves represent disintegration mechanisms.}
    \label{fig:coll_d_14}
\end{figure}

\begin{figure}
    \centering
    \includegraphics[width=\columnwidth]{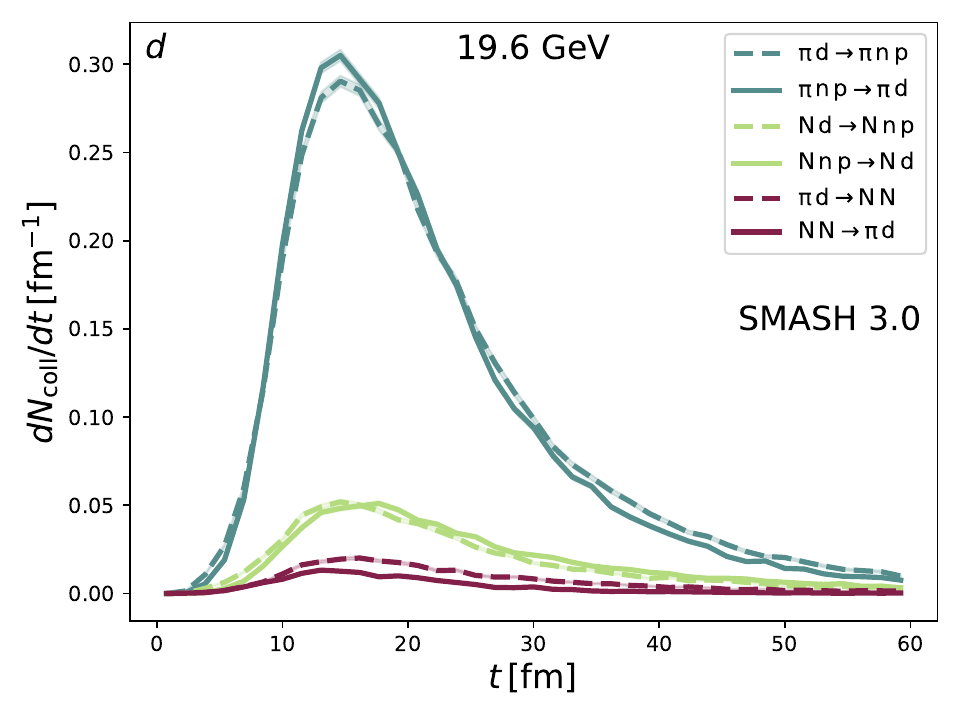}
    \caption{Scattering rates for deuterons at 19.6 GeV in central Au-Au-collisions. The solid curves represent formation mechanisms while the dashed curves represent disintegration mechanisms.}
    \label{fig:coll_d_19}
\end{figure}

\begin{figure}
    \centering
    \includegraphics[width=\columnwidth]{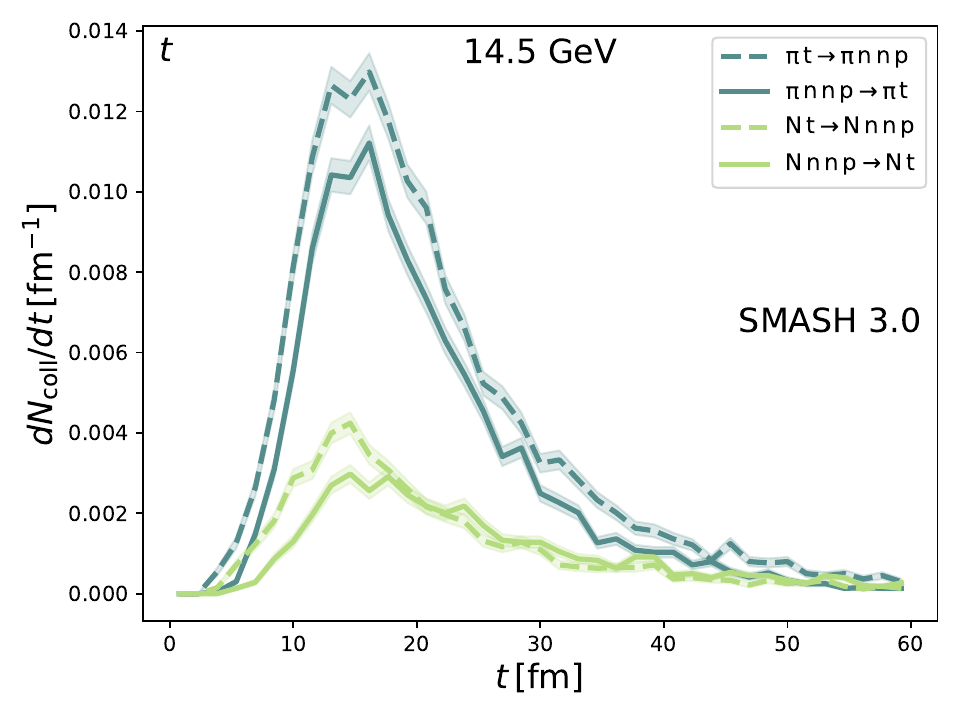}
    \caption{Scattering rates for tritons at 14.5 GeV in central Au-Au-collisions. The solid curves represent formation mechanisms while the dashed curves represent disintegration mechanisms.}
    \label{fig:coll_t_14}
\end{figure}

\begin{figure}
    \centering
    \includegraphics[width=\columnwidth]{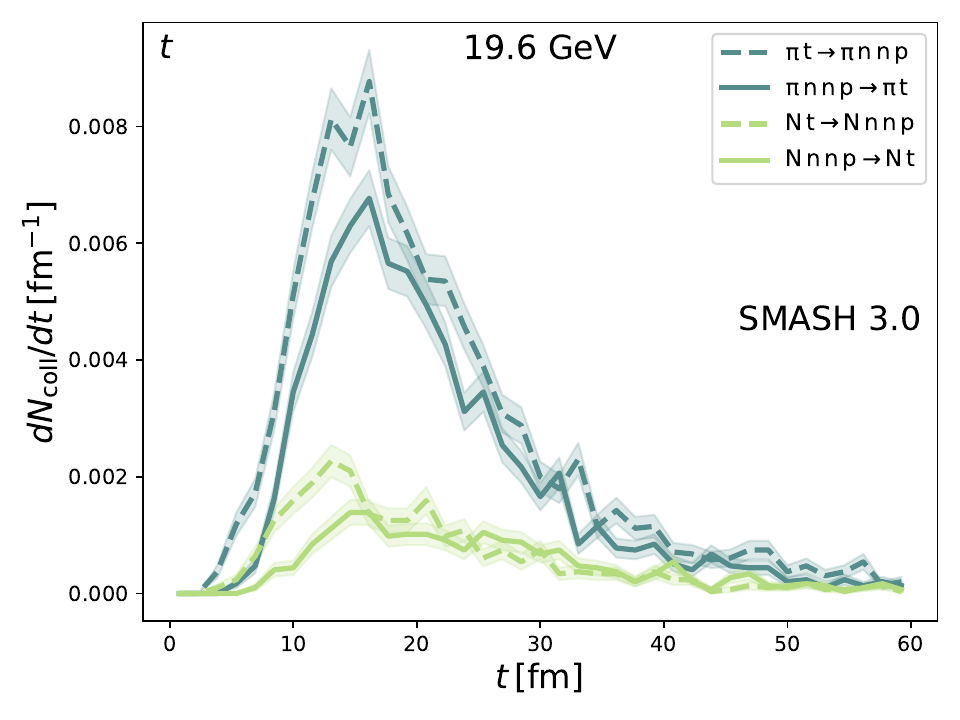}
    \caption{Scattering rates for tritons at 19.6 GeV in central Au-Au-collisions. The solid curves represent formation mechanisms while the dashed curves represent disintegration mechanisms.}
    \label{fig:coll_t_19}
\end{figure}

\begin{figure}
    \centering
    \includegraphics[width=\columnwidth]{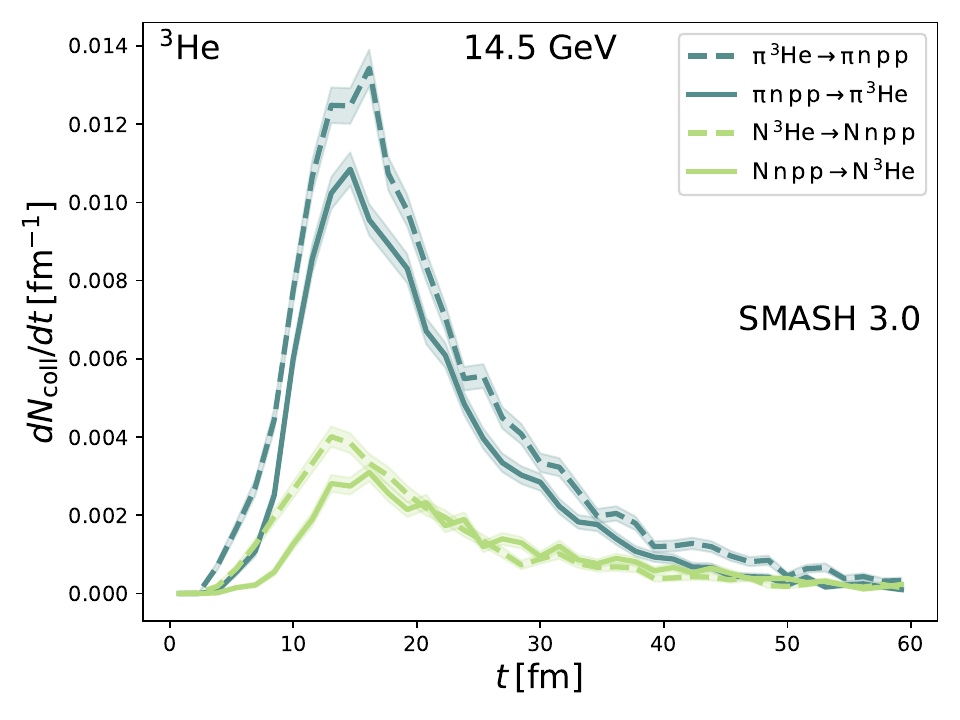}
    \caption{Scattering rates for $^3\mathrm{He}$ nuclei at 14.5 GeV in central Au-Au-collisions. The solid curves represent formation mechanisms while the dashed curves represent disintegration mechanisms.}
    \label{fig:coll_h_14}
\end{figure}

\begin{figure}
    \centering
    \includegraphics[width=\columnwidth]{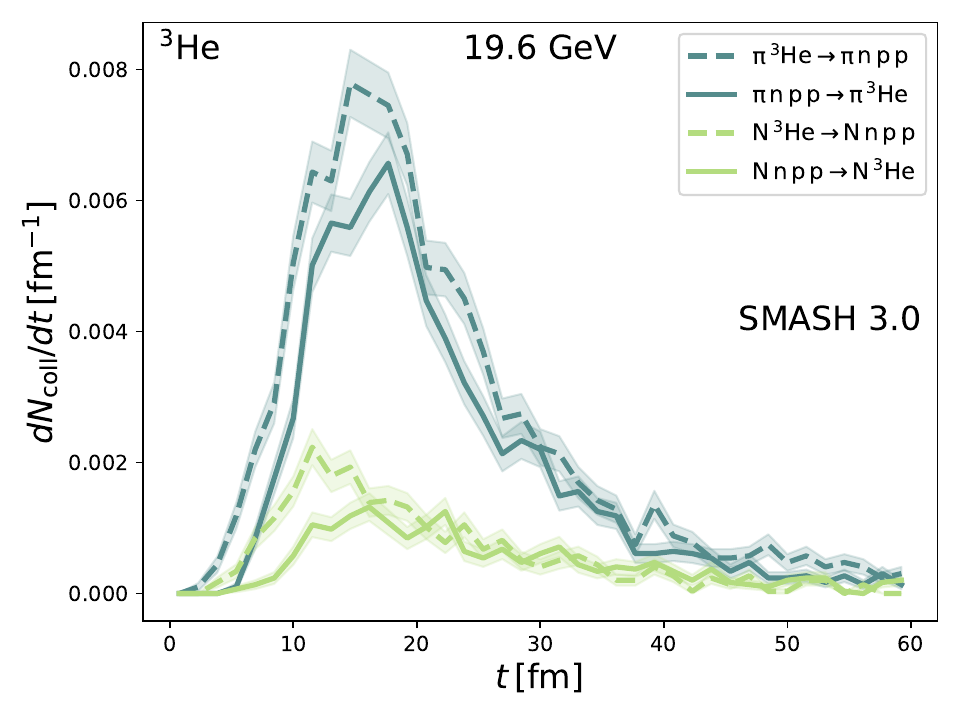}
    \caption{Scattering rates for $^3\mathrm{He}$ nuclei at 19.6 GeV in central Au-Au-collisions. The solid curves represent formation mechanisms while the dashed curves represent disintegration mechanisms.}
    \label{fig:coll_h_19}
\end{figure}

\begin{figure}
    \centering
    \includegraphics[width=\columnwidth]{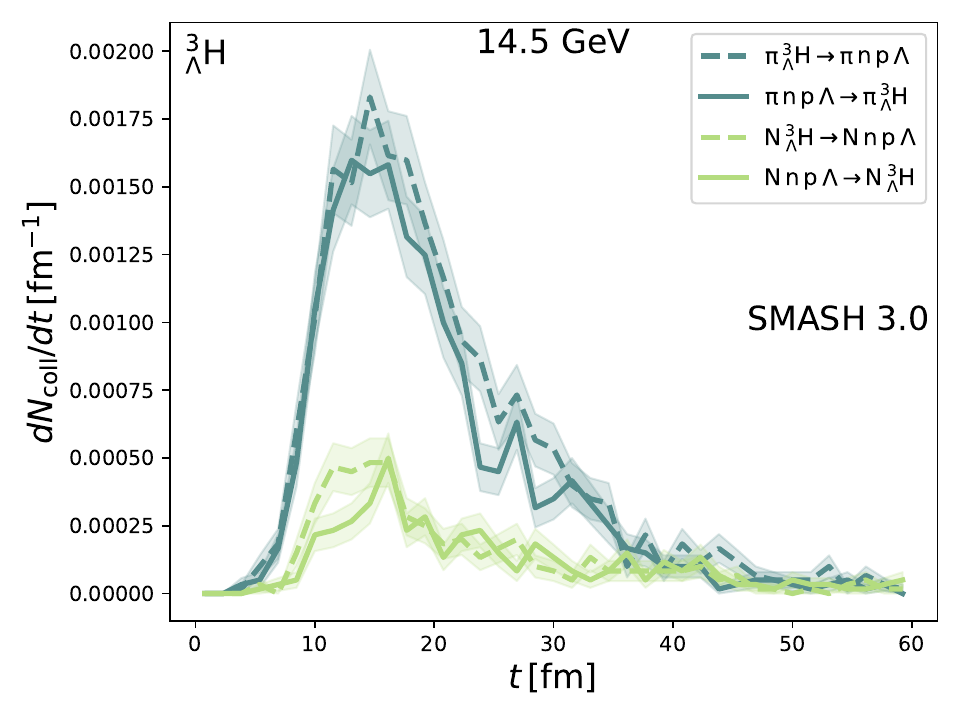}
    \caption{Scattering rates for hypertritons at 14.5 GeV in central Au-Au-collisions. The solid curves represent formation mechanisms while the dashed curves represent disintegration mechanisms.}
    \label{fig:coll_hy_14}
\end{figure}

\begin{figure}
    \centering
    \includegraphics[width=\columnwidth]{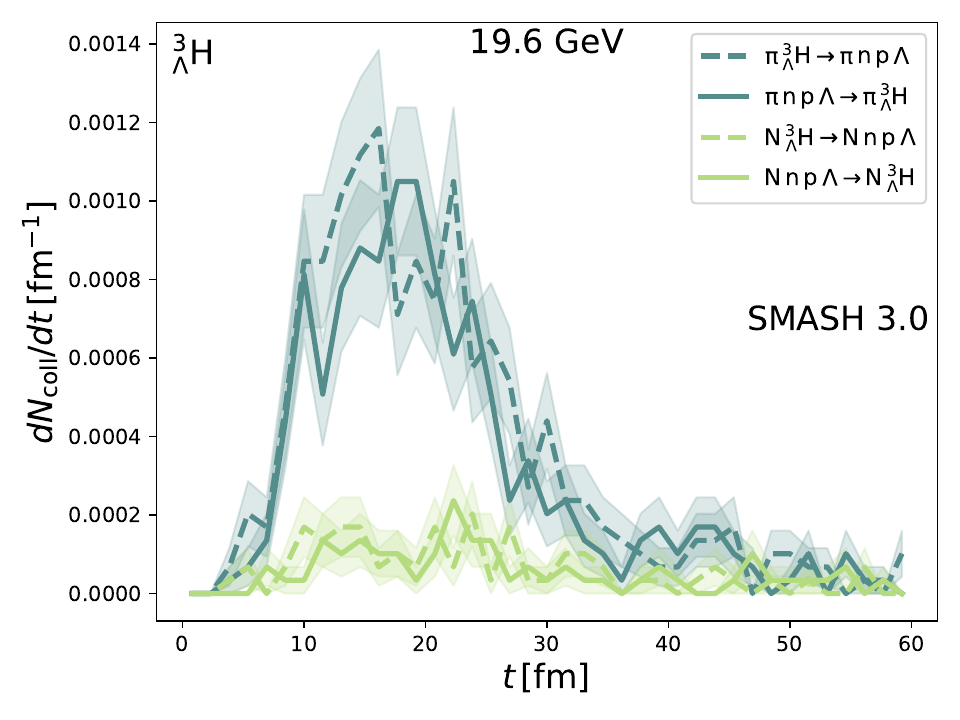}
    \caption{Scattering rates for hypertritons at 19.6 GeV in central Au-Au-collisions. The solid curves represent formation mechanisms while the dashed curves represent disintegration mechanisms.}
    \label{fig:coll_hy_19}
\end{figure}

In Section \ref{sec:productionmech} we only showed the results for the lowest energy 7.7 GeV.
Additionally we present here the results for 14.5 GeV and 19.6 GeV as well.
Figures \ref{fig:coll_d_14}-\ref{fig:coll_hy_19} show the scattering rates for deuterons, tritons, $^3\mathrm{He}$ nuclei and hypertritons.
Nuclei formation via pion (blue curve) or nucleon (green curve) catalysis reactions is compared to disintegration.
As already discussed in Section \ref{sec:productionmech}, deuterons can also be formed/disintegrated via inelastic $2\rightarrow2$ scattering as depicted by the red curves in Figures \ref{fig:coll_d_14} and \ref{fig:coll_d_19}.

\clearpage
\bibliography{str_triton}% Produces the bibliography via BibTeX.

\end{document}